\documentclass[%
 reprint,
superscriptaddress,
 amsmath,amssymb
 aps,floatfix]{revtex4-2}

\usepackage[utf8]{inputenc}
\usepackage[T1]{fontenc}
\usepackage{blindtext}
\usepackage{natbib}
\usepackage{graphicx}% Include figure files
\usepackage{dcolumn}% Align table columns on decimal point
\usepackage{bm}% bold math
\usepackage{hyperref}% add hypertext capabilities
\usepackage[norsk,nameinlink]{cleveref}
\usepackage{siunitx}
\usepackage{xspace}
\usepackage{xcolor}
\usepackage{multirow}
\usepackage{ulem}
\usepackage{microtype}
\Crefname{figure}{Fig.}{Figs.}
\Crefname{table}{\text{Table}}{\text{Table}}
\Crefname{equation}{\text{Eq.}}{\text{Eqs.}}
\Crefname{section}{\text{section}}{sections}

\Crefname{subsection}{subsection}{subsections}
\Crefname{appendix}{Appendix}{Appendices}

\newcommand{\SiOO}{SiO$_{2}$\xspace}

\newcommand{\LEmuSR}{LE-$\mu$SR\xspace}
\newcommand{\muSR}{$\mu$SR\xspace}

\newcommand{\FD}{F$_\textrm{D}$\xspace}
\newcommand{\FMu}{F$_\textrm{Mu}$\xspace}

\newcommand{\mup}{$\mu^+$\xspace}

\newcommand{\Mum}{Mu$^-$\xspace}
\newcommand{\Mup}{Mu$^+$\xspace}
\newcommand{\Muz}{Mu$^0$\xspace}
\newcommand{\VC}{V$_\textrm{C}$\xspace}
\newcommand{\VSi}{V$_\textrm{Si}$\xspace}
\newcommand{\Muone}{Mu$_1$\xspace}
\newcommand{\Mutwo}{Mu$_2$\xspace}
\newcommand{\ND}{N$_D$\xspace}
\newcommand{\WD}{$w_D$\xspace}
\newcommand{\EA}{$E_\text{A}$\xspace}

\begin{document}

\preprint{APS/123-QED}

\title{Muonium dynamics as a probe for depth-resolved properties of 4H-SiC}% Force line breaks with \\

 \author{Maria Mendes Martins}%
 \email[]{martins@aps.ee.ethz.ch}
 \affiliation{PSI Center for Neutron and Muon Sciences, 5232 Villigen PSI, Switzerland}
\affiliation{Advanced Power Semiconductor Laboratory, ETH Zurich, 8092 Zurich, Switzerland}

 \author{Piyush Kumar}
  
  \affiliation{Advanced Power Semiconductor Laboratory, ETH Zurich, 8092 Zurich, Switzerland}


 \author{Marianne E. Bathen}
 \affiliation{Advanced Power Semiconductor Laboratory, ETH Zurich, 8092 Zurich, Switzerland}
 \affiliation{
  Department of Physics/ Centre for Materials Science and Nanotechnology, University of Oslo, 0316 Oslo, Norway}

\author{Robert J. Scheuermann}
\affiliation{PSI Center for Neutron and Muon Sciences, 5232 Villigen PSI, Switzerland}

\author{Zurab Guguchia}
\affiliation{PSI Center for Neutron and Muon Sciences, 5232 Villigen PSI, Switzerland}

\author{Lasse Vines}
 \affiliation{
  Department of Physics/ Centre for Materials Science and Nanotechnology, University of Oslo, 0316 Oslo, Norway}

 \author{Ulrike Grossner}
  \affiliation{Advanced Power Semiconductor Laboratory, ETH Zurich, 8092 Zurich, Switzerland}
\author{Thomas Prokscha}
 \email[]{thomas.prokscha@psi.ch}
\affiliation{PSI Center for Neutron and Muon Sciences, 5232 Villigen PSI, Switzerland}

\date{\today}

\begin{abstract}

This study establishes a baseline for muonium (Mu) charge-exchange dynamics in n-type 4H-SiC through a detailed low-energy muon spin rotation (\LEmuSR) investigation. Epitaxially grown and ion-implanted samples with nitrogen and phosphorus donors were characterized to assess the effect of carrier concentration and doping method on defect formation. \LEmuSR enabled nanometer-scale depth profiling of near-surface and implanted regions, revealing variations in charge carrier concentration due to fixed surface charges. The temperature dependence of the diamagnetic fraction and phase provided direct evidence of the \Muz to \Mum transition, with extracted activation energies consistent with known donor ionization energies. Additionally, high-field \muSR was used to analyze the Mu dynamics, and Monte-Carlo simulations to model the \Muz electron capture process. The simulation results offer a quantitative method to extract free electron concentrations from \LEmuSR data, enhancing its capability to characterize the activation of dopants and carrier depth profiles. 
We demonstrate that \LEmuSR is a powerful depth-resolved tool that can provide insights for optimizing the fabrication of reliable SiC devices for power electronics.
\end{abstract}

\maketitle

\section{\label{sec:level1}Introduction}

Silicon carbide (4H-SiC) is a wide-bandgap semiconductor that is increasingly adopted due to its suitability for high-temperature, high-power, and high-frequency applications.
One of the key technological developments allowing the manufacturing of reliable devices is the epitaxial growth of 4H-SiC, which enables the formation of layers with precise control over their thickness, doping concentrations, and relatively low defect densities \cite{Choyke_2004, Kimoto_2014Fund}.
Selective doping is another essential step in device manufacturing. However, due to the strong silicon-carbon (Si–C) bonds in the crystal lattice, the dopant diffusivity is very low. As a result, ion implantation is the preferred method for introducing dopants into specific regions of the semiconductor material. While this technique allows for accurate control over doping profiles, it also introduces significant lattice damage that must be carefully managed \cite{Pensl_2003}.
For n-type doping, nitrogen (N) and phosphorus (P) are commonly used, while p-type doping is typically achieved using aluminum (Al), and less frequently, boron (B) \cite{Kimoto_2014, Roccaforte_2021}.

A post-implantation annealing (PIA) step is required to recover the structure and electrically activate the dopants, by promoting their inclusion into the SiC lattice. Annealing between \SIlist{1600;1700}{\celsius} results in almost \SI{100}{\percent} electrical activation of N and P donors \cite{Blanque_2004,Kimoto_2014}. During the high-temperature annealing, the samples are usually coated with a carbon (C) cap to prevent Si desorption and surface degradation \cite{Negoro_2004,Vassilevski_2005}. The presence of a C-cap allows carbon to be injected into SiC, contributing to the reduction of the density of carbon vacancies (V$_\mathrm{C}$) \cite{Ayedh_2017thermodynamic} and the formation of other carbon-related defects \cite{Karsthof_2022Cint,Bathen2024Cinj}.

Deep-level transient spectroscopy (DLTS) is commonly used in the study of defects in the bulk of semiconductors, since electrically active defects and their location within the bandgap can be identified. DLTS measurements have shown that even after PIA, some of the defects created during ion implantation are still present and can extend beyond the projected implantation profiles \cite{Kawahara_2010}.
However, the DLTS probing depth and sensitivity to defect densities are limited by the doping concentration, and it is hard to obtain information about the near-surface region (top hundred nanometers). Therefore, low-energy muon spin rotation spectroscopy (\LEmuSR) \cite{Morenzoni_2000,Bakule_2004,Prokscha_2008} was used to study the effect of N and P ion implantation by directly probing the implanted regions from the surface to a depth of \SI{150}{\nano\meter}, while varying the muons' implantation energy. 
\LEmuSR studies of 4H-SiC have allowed meaningful insight into the properties of defects that affect the reliability of devices (\VC) and ones with applications in quantum technologies (\VSi) \cite{Woerle_2019a,Woerle_2020}. Oxide-semiconductor systems were investigated, and the effects of oxidation and post-annealing on the creation and passivation of defects created on the interface and near-interface could be determined \cite{Martins_2023,Kumar_2023,Kumar_2024_AlImp}. The muon is also very sensitive to the charge carriers in its vicinity \cite{Patterson_1988} and has been used to quantify changes in the electrical environment as a function of depth with nanometer depth resolution \cite{Prokscha_2007, Prokscha_2020}.

We present the high-field \muSR measurements performed using the High-field and Low temperature instrument (HAL-9500), where the temperature dependence of the \muSR signal is used to characterize the different muonium (Mu) states formed in SiC and transitions between them. A good understanding of the Mu dynamics in the material, e.g., activation energy of the transition, lattice sites and charge states occupied, is crucial to use the muon as a probe for the processing-induced defects.

Previous \muSR works \cite{Lichti_2004,Celebi_2009} show the presence of two distinct \Muz centers in the high field regime (B>\SI{0.5}{\tesla}) with hyperfine constants A$_\textrm{HF}$ = \SI{3030}{MHz} and \SI{2825}{MHz}, called \Muone and \Mutwo, respectively. These states are observed in n-type, p-type and high resistivity 4H-SiC. However, important details about the samples studied with \muSR, e.g. doping concentration, growth process, were not reported.
More recently, \LEmuSR measurements at low magnetic field reveal a previously unreported weak anisotropy of the hyperfine coupling \cite{Martins_2023}. This is seen as a splitting in the \Muz precession frequency into two lines, which stays constant in the measured field range.

The sensitivity of the muon probe to defects and charge carrier density, combined with its nanometer depth resolution, means that \LEmuSR can provide information about the activation of donors in n-type 4H-SiC, and reveal differences between the doping process during epitaxial growth and ion-implantation, as well as the different defects created due to N- vs P-implantation. By using \LEmuSR, the implantation region and the near-surface are directly investigated. 
This investigation is an important building block in the characterization of 4H-SiC interfaces and devices, providing a baseline of the \LEmuSR sensitivity to the charge carrier concentration.

\noindent

\section{Methodology}

\subsection{Sample Preparation}
\begin{figure}[hbt!]

\includegraphics[width=0.4\textwidth]{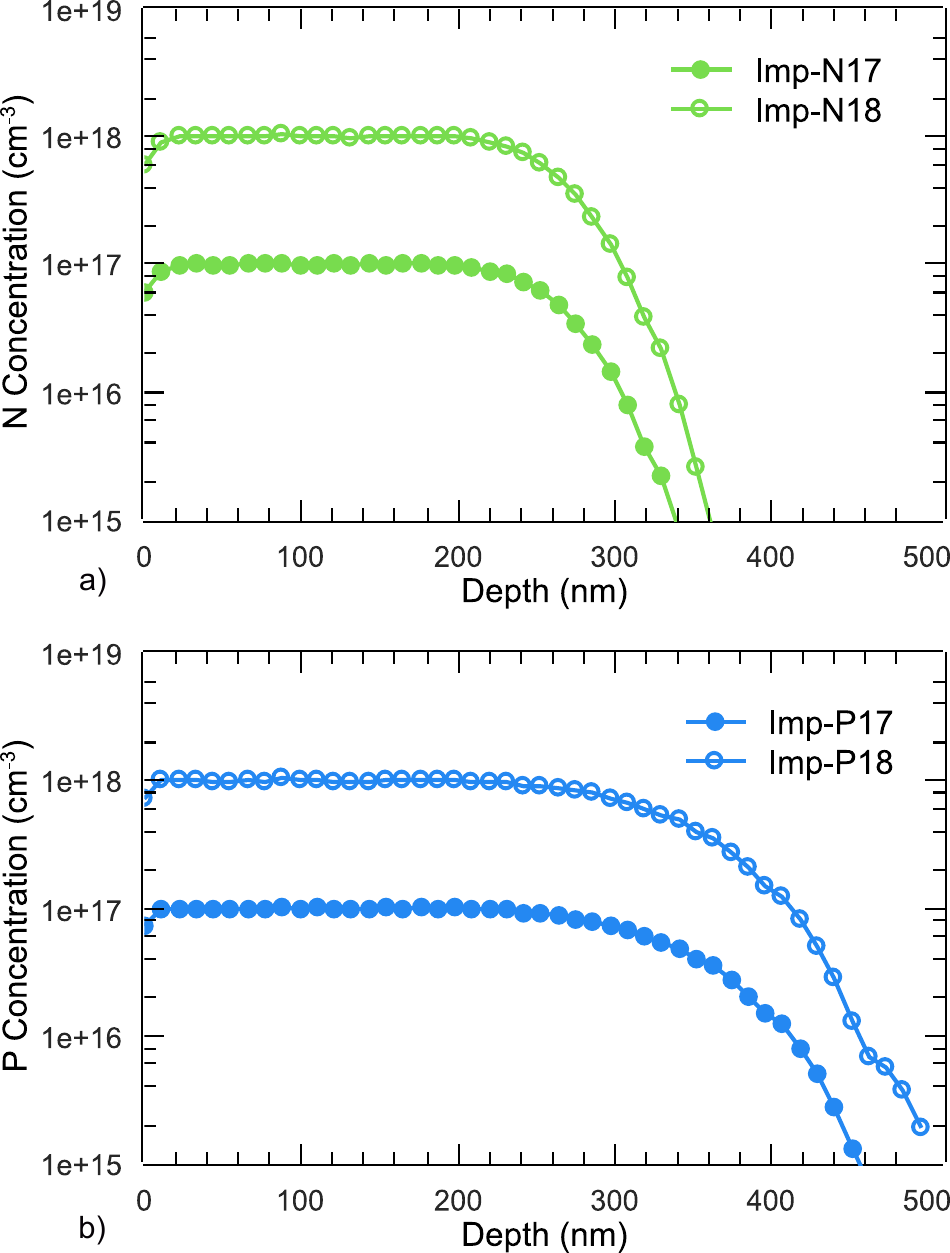}
\caption{\label{fig:SRIMImp-N-P} Box profiles for the implantation of a) N$^+$ and b) P$^+$ ions in the 4H-SiC epilayer samples. The ion-implantation profiles were obtained with SRIM simulations \cite{Ziegler_1985,Ziegler_2010}.}
\end{figure}

For this study, n-type 4H-SiC epilayers with a thickness of \SI{10}{\micro\meter} grown on highly-doped 4H-SiC substrates were characterized with \LEmuSR. The epilayer is a high-purity single crystal, grown on substrates that are \SI{4}{\degree} off the (0001) c-axis, maintaining the 4H stacking sequence. The details on sample preparation are found in \Cref{tab:Samples}. The epitaxial samples were doped with nitrogen during growth, and have doping concentrations of \SIlist{4e15;1e17}{\per\cubic \centi\meter}. The implanted samples were prepared on n-type epilayers with N$_D$=\SI{3e15}{\per\cubic\centi\meter} background doping. In order to create regions with well defined doping concentration with thicknesses between \SIlist[list-units=single]{250;350}{\nano\meter}, multiple ion (N or P) implants at different energies and fluences were performed at the Fraunhofer IISB in Erlangen, Germany. The resulting implantation box profiles are shown in \Cref{fig:SRIMImp-N-P}. The N-implantation consisted of five implantation steps performed with energies between \SIrange[range-units=single]{200}{400}{\kilo\electronvolt}, and the P-implantation of six steps with energies between \SIrange[range-units=single]{300}{800}{\kilo\electronvolt}.
The total fluence of the implantation is \SI{4e12}{\per\square\centi\meter} and \SI{4e13}{\per\square\centi\meter} for the doping concentrations \SI{2e17}{\per\cubic\centi\meter} and \SI{1e18}{\per\cubic\centi\meter}, respectively.
Before implantation, \SI{550}{\nano\meter} of \SiOO was deposited using plasma-enhanced physical vapor deposition (PECVD). The sacrificial \SiOO layer is used to guarantee the implantation peaks of the ions are in the SiC epilayer. Implantation was performed with the samples held at $\sim$\SI{500}{\celsius} to increase dynamic annealing of the lattice damage simultaneously with the formation of ion damage \cite{kuznetsov_dynamic_2003}. After implantation, the deposited oxide was removed using hydrofluoric (HF) acid. Finally, the samples were cleaned and coated with a carbon cap (C-cap), to be annealed at \SI{1650}{\celsius} for 30~minutes. After annealing, the carbon cap was removed using an oxygen plasma, and subsequently cleaned in HF.

An additional n-type 4H-SiC sample (N-doped), with a \SI{150}{\micro\meter} thick epilayer with N$_D=$~\SI{2e14}{\per\cubic\centi\meter} grown on a \SI{350}{\micro\meter} substrate and purchased from Ascatron, was investigated in the HAL-9500 instrument, which uses a conventional \SI{4.1}{\mega\electronvolt} surface muon beam to study the bulk of the material.

\begin{table*}[h!t]
\caption{\label{tab:Samples} Description of the 4H-SiC samples investigated with \LEmuSR. The parameters are the doping concentration N$_D$, donor species, doping process, and post-implantation annealing (PIA) with a carbon capping layer. Implantation was performed to form the box profiles shown in \Cref{fig:SRIMImp-N-P}. N$_D$ was extracted from capacitance-voltage measurements. }
\begin{ruledtabular}
\begin{tabular}{ccccc}
\textbf{Sample name} & \textbf{N$_D$ (\SI{}{\per\cubic\centi\meter})} &\textbf{ Donor Species} & \textbf{Doping process} & \textbf{PIA parameters}\\
\hline

Epi-N15 & \SI{4e15}{} & \multirow{2}{*}{Nitrogen} &  \multirow{2}{*}{Epitaxial growth}  & -- \\
Epi-N17 & \SI{1e17}{} &   &  & --\\ \hline 
Imp-N17  &\SI{2e17}{} & \multirow{2}{*}{Nitrogen}  & \multirow{4}{*}{Ion Implantation} &  \multirow{4}{*}{30 minutes at \SI{1650}{\celsius}}\\ 
Imp-N18  &\SI{1e18 }{} &  & & \\ \cline{1-3}
Imp-P17  & \SI{2e17}{} & \multirow{2}{*}{Phosphorus} &  & \\
Imp-P18  & \SI{1e18}{} &   &  & \\

\end{tabular}
\end{ruledtabular}
\end{table*}

\subsection{Muon spin spectroscopy}
The muon spin rotation technique (\muSR) uses the positive muon \mup as a local probe to study the electrical and magnetic properties in solids \cite{Blundell_1999,Blundell_2021_muon_spectroscopy,Hillier_2022_primer,Amato_introduction_2024}. A beam of nearly \SI{100}{\percent} spin-polarized muons is implanted into the sample under study. 
The implanted \mup decays with a mean lifetime of $\tau_\mu \approx $~\SI{2.2}{\micro \second}
and spontaneously emits a positron preferentially in the direction of the muon spin at the time of the decay. 

By detecting the emitted positron, the time evolution of the muon spin polarization in the sample is obtained.  In the sample, the muon comes to rest at an interstitial site of the host lattice and can be treated as a light isotope of hydrogen called muonium (Mu). In semiconductors and insulators, the \mup probe can stay positive in the \Mup state, capture an electron and form the paramagnetic state \Muz, or capture even two electrons to form the negative \Mum state, if enough electrons are available in its surroundings. The charged states \Mup and \Mum without coupling to an unpaired electron are called diamagnetic. In the measurements, we used a transverse-field configuration (TF-\muSR), where an external magnetic field was applied perpendicular to the spin of the incoming muon, inducing precession of the muon spin. In the case of the two diamagnetic Mu states, the muon spin precesses at its Larmor frequency, whereas the hyperfine coupling with the unpaired electron in the \Muz state causes different muon spin precession frequencies, which are typically very distinct from the Larmor frequency \cite{Patterson_1988}. The muon spin precession frequencies corresponding to the transitions between the four hyperfine levels of the \Muz state can be detected, and information about the occupancy and dynamics of the diamagnetic and paramagnetic states can be extracted. \\

\begin{figure}[hbt!]
\includegraphics[width=0.4\textwidth]{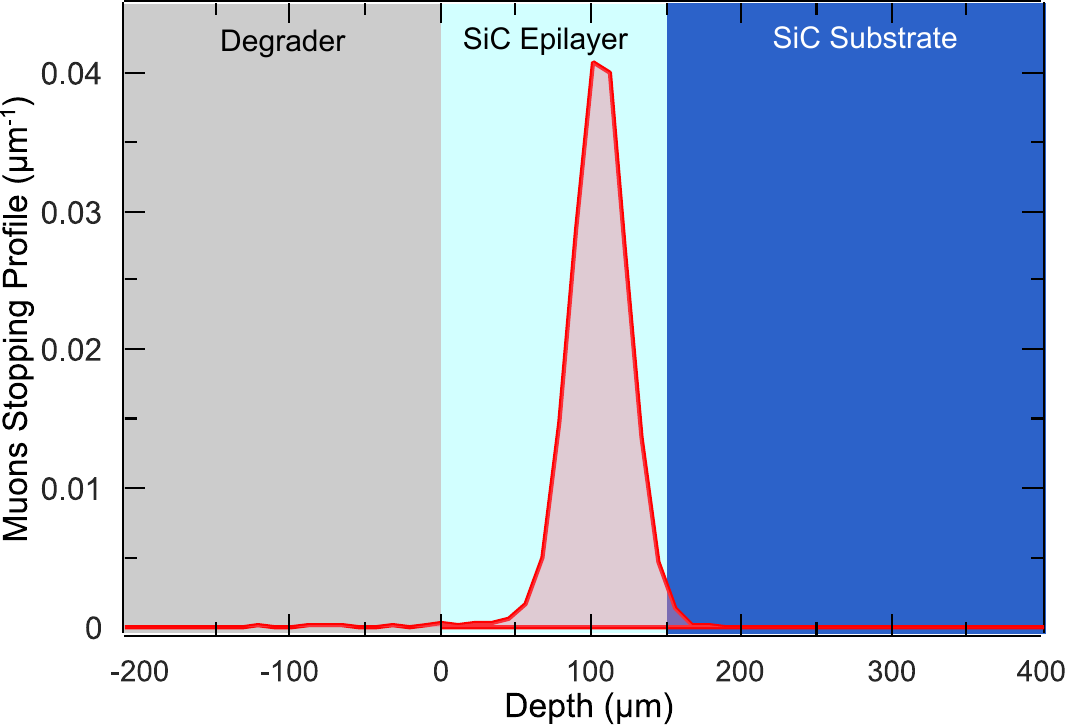}
\caption{\label{fig:stopProfile_SRIM} Stopping distribution of the muon beam calculated with SRIM simulation \cite{Ziegler_1985,Ziegler_2010}. The curve shows that the majority of the muons stop in the \SI{150}{\micro\meter} thick SiC epilayer, with a small fraction of the muons stopping in the highly doped substrate layer.}
\end{figure}

Typically, \mup beams with an energy of \SI{4.1}{\mega\electronvolt} are available at proton accelerator facilities.
At this energy, the muons penetrate several hundred micrometers deep and the bulk of the material is probed. We performed high-field \muSR measurements in HAL-9500 under a high magnetic field of \SI{6}{\tesla} to identify the TF precession frequencies of the muonium states in low-doped 4H-SiC and to compare with previous \muSR experiments \cite{Lichti_2004}. 
To stop the incoming muon beam in the low-doped \SI{150}{\micro\meter} epilayer, the muons were slowed down by using a degrader. The degrader was chosen by simulating the stopping of the muons with SRIM \cite{Ziegler_1985,Ziegler_2010} (\Cref{fig:stopProfile_SRIM}), and consisted of a stack of \SI{200}{\micro\meter} Al, \SI{10}{\micro\meter} Ag and \SI{25}{\micro\meter} Al.

\subsection{Low-energy \muSR}
\begin{figure}[hbt!]

\includegraphics[width=0.4\textwidth]{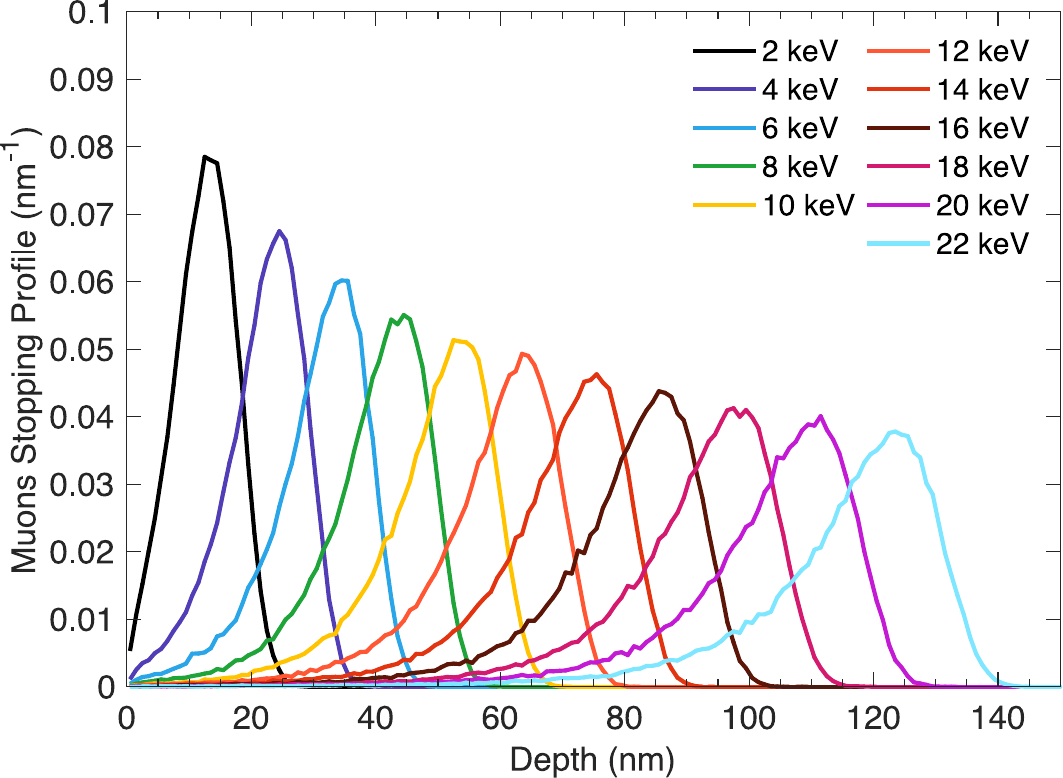}
\caption{\label{fig:stopProfiles} Stopping profiles of muons for the different implantation energies, obtained for 4H-SiC with the Monte-Carlo simulation program TRIM.SP \cite{Morenzoni_2002, Eckstein_1991}.}
\end{figure}

\LEmuSR is an extension of \muSR with interesting applications in the study of thin-films, surfaces, and interfaces, due to its nanometer depth resolution. The probing depth can be varied between the near-surface to around \SI{150}{\nano\meter} in 4H-SiC \cite{Morenzoni_2000,Bakule_2004}. The stopping profiles of the muon beam for different implantation energies are shown in \Cref{fig:stopProfiles}, and are simulated using the Monte-Carlo simulation program TRIM.SP \cite{Morenzoni_2002,Eckstein_1991}.

The \LEmuSR measurements were performed at TF values of \SIlist[list-units=single]{0.7;10}{\milli\tesla}. It is necessary to collect data at low magnetic fields (B$<$\SI{1}{\milli\tesla}) to observe the muonium precession frequency, due to the upper detection limit of about \SI{60}{\mega\hertz} of the low-energy muon spectrometer (LEM), which stems from the limited time resolution of LEM of about \SI{5}{\nano\second}. 
%During the low magnetic field measurements, the field was set to \SI{0.7}{\milli\tesla}; however, the fits of the data (detailed below) reveal the external field to be \SI{0.76}{\milli\tesla}. 
%At \SI{0.7}{\milli\tesla}, the diamagnetic and about \SI{50}{\percent} of one of the paramagnetic components can be spectroscopically detected.  
\textcolor{black}{At \SI{0.7}{\milli\tesla}, the diamagnetic and the \Muone paramagnetic component can be spectroscopically detected, whilst the \Mutwo paramagnetic component depolarizes too quickly to be observed.}
At this low field, the transition frequencies between the triplet states in Mu$^0$ with isotropic hyperfine coupling $\nu_{12}$ and $\nu_{23}$ overlap, resulting in a single Mu$^0$ precession frequency. However, a splitting of this precession frequency, due to a weak anisotropy of the hyperfine coupling \cite{Lichti_2004,Martins_2023}, results in the spectrum shown in \Cref{fig:low-dopedSiCMu_LEM}~a). At \SI{10}{\milli\tesla}, the \Muz transition frequencies are higher than the maximum resolution of the
LEM spectrometer,
thus only the diamagnetic component is observed and analyzed.

The temperature dependence was measured in HAL-9500 and LEM between \SIlist{10;310}{\kelvin} to observe the temperature-activated transitions of the Mu states before and after the charge carriers were free in 4H-SiC. The energy dependence was measured in LEM at \SI{260}{\kelvin}, when the activation of the donors is approximately \SI{100}{\percent}, to probe changes with depth in the epilayer and implanted region. The muon implantation energy was varied between \SIlist[list-units=single]{2;22}{\kilo\electronvolt}, corresponding to a mean probing range of \SIlist[list-units=single]{5;130}{\nano\meter} (\Cref{fig:stopProfiles}), respectively.

\subsection{Data analysis}
The \muSR data analysis was performed using the musrfit software \cite{Suter_2012}. In HAL-9500, at a transverse magnetic field of \SI{6}{\tesla}, five lines are observed in the Fourier spectrum (see \Cref{fig:HAL-Tscan}~a)) --- one diamagnetic line and four lines resulting from the formation of two different muonium configurations in 4H-SiC (\Muone and \Mutwo). At \SI{6}{\tesla}, two precession frequencies are observed for each Mu state, $\nu_{12}$ and $\nu_{34}$. The data obtained from HAL-9500 were thus fitted with five components as follows:
\begin{equation}
\begin{split}
    A(t) &= A_\text{D} \cdot \exp(-\lambda_\text{D} \cdot t) \cdot \cos(2\pi\nu_D t + \phi_\text{D}) \\
       &+  A_{\nu_{12}}^\text{Mu1} \cdot \exp(-\lambda_{\nu_{12}}^\text{Mu1}\cdot t)\cdot \cos(2\pi\nu_{\nu_{12}}^\text{Mu1} t + \phi_{\nu_{12}}^\text{Mu1}) \\&+  A_{\nu_{34}}^\text{Mu1} \cdot \exp(-\lambda_{\nu_{34}}^\text{Mu1}\cdot t)\cdot \cos(2\pi\nu_{\nu_{34}}^\text{Mu1} t + \phi_{\nu_{34}}^\text{Mu2}) \\&+
       A_{\nu_{12}}^\text{Mu2} \cdot \exp(-\lambda_{\nu_{12}}^\text{Mu2}\cdot t)\cdot \cos(2\pi\nu_{\nu_{12}}^\text{Mu2} t + \phi_{\nu_{12}}^\text{Mu2}) \\&+  A_{\nu_{34}}^\text{Mu2} \cdot \exp(-\lambda_{\nu_{34}}^\text{Mu2}\cdot t)\cdot \cos(2\pi\nu_{\nu_{34}}^\text{Mu2} t + \phi_{\nu_{34}}^\text{Mu2}).
\end{split}
\label{eq:Lorentzian5C}
\end{equation}

The observed amplitude of the high-frequency lines is reduced due to the finite time resolution. Thus, all the amplitudes, including the Ag calibration, were corrected using the correction function $C(\nu)$ \cite{holzschuh_direct_1981}:
\begin{equation}
    C(\nu)=\exp[-(2\pi\nu\sigma)^2/2],
\end{equation}
where the width is $\sigma=\SI{66}{\pico\second}$, calculated as:
\begin{equation}
    \sigma=\frac{1}{\pi\sqrt{2}}\cdot \sqrt{ \ln\frac{A_{\nu12}}{A_{\nu34}}\left( \nu_{34}^2 - \nu_{12}^2  \right)}.
\end{equation}

Each component was fitted with its own relaxation rate $\lambda$. For each \Muz, $\lambda_{\nu_{12}}$ and $\lambda_{\nu_{34}}$ values are similar, thus the average $\lambda_\text{Mu1}$, and $\lambda_\text{Mu2}$ were calculated.

The \LEmuSR time spectra in \Cref{fig:low-dopedSiCMu_LEM}~a) were fitted with a three-component function of the form:
\begin{equation}
  \begin{split}
       A(t) &= A_\text{D} \cdot \exp(-\lambda_\text{D} \cdot t) \cdot \cos(2\pi\nu_D t + \phi_\text{D})\\
       &+ A_\text{Mu} \cdot \exp(-\lambda_\text{Mu}\cdot t)\cdot 
          [\cos(2\pi\nu_1 t + \phi_\text{Mu}) \\
       &+ \cos(2\pi\nu_2 t + \phi_\text{Mu})],
  \end{split}
     \label{eq:Lorentzian2C}
\end{equation}

where $A_\text{D}$ is the asymmetry of the diamagnetic component (either \Mup or \Mum), $\lambda_\text{Mu}$ is the exponential depolarization rate of \Muz,
$\nu_{1/2} = \nu_\text{Mu}\pm\nu_a/2$ with the low-field muon spin precession frequency $\nu_\text{Mu}$ in the \Muz triplet state without anisotropy, and $\nu_a = 1.7$~MHz is the small anisotropic component of the hyperfine coupling on top of the isotropic part of $\sim 3000$~MHz \cite{Martins_2023}.
The phase $\phi_\textrm{D}$ of the diamagnetic signal is usually the angle offset of the decay positron detector in relation to the muon spin polarization at $t=0$. However, a possible phase shift, observed as a function of temperature, indicates the occurrence of conversion between muonium states --- due to a change of the muon spin precession frequencies \cite{Patterson_1988}. The data collected at \SI{10}{\milli\tesla} was fitted using only the first component of \Cref{eq:Lorentzian2C}, since only the diamagnetic contribution is observable in LEM at this field.

The asymmetry parameters  $A_\text{D}$ and  $A_\text{Mu}$ were converted to the corresponding fractions \FD and \FMu, by normalizing to the maximum asymmetry measurable in silver ($A_\text{Ag}$). The fractions are calculated as \FD$=A_\text{D}/A_\text{Ag}$ and \FMu$=2\cdot A_\text{Mu}/A_\text{Ag}$. The factor of 2 in \FMu accounts for the fact that only \SI{50}{\percent} of the \Muz polarization is observable in low fields \cite{Blundell_2021_muon_spectroscopy}.
 
Importantly, deconvolution of the experimental data as a function of implantation energy and the stopping depth profile of the muon is performed to model the diamagnetic fraction as a function of sample depth, as detailed in Ref.~\cite{martins_depth_2023}. The variation of \FD with depth using a multilayer fitting approach allows observing local changes in carrier concentration, and to extract the charge carrier depth profiles.

\noindent

\subsection{\label{subsec:simulation}Monte-Carlo simulation of the electron capture}
In semiconductors, transitions between the different muonium states may occur, depending on the temperature, concentration, and type of the free carriers present in the sample \cite{Patterson_1988,Prokscha_2020,Hitti_1999}.
In n-type SiC, the electron capture process \cite{BaniSalameh_2007}
\begin{equation*}
    \text{Mu$^0$} + \text{e}^- \rightarrow \text{Mu$^-$}
\end{equation*} occurs  when the dopants are ionized (T$>\SI{50}{\kelvin}$) \cite{kimoto_nitrogen_1995} and is governed by the capture rate 
\begin{equation}\label{eq:lambda_fit}
    \lambda_\text{c}=n \cdot v_e \cdot \sigma^{0/-},
\end{equation}

where $n$ is the effective free electron concentration, $v_e$ is the thermal velocity of the electrons,
and $\sigma^{0/-}$ is the electron capture cross-section for this process. Note, that all these parameters are a function of temperature, as discussed in \Cref{sec:Appendix_ElectronCapture}.

The impact of the muonium dynamics in 4H-SiC was modeled using a Monte-Carlo simulation \cite{Prokscha_2012,Prokscha_2014}, by calculating the temporal progression of the muon spin phase in a given magnetic field. 
In this work, only the electron capture process at \SIlist[list-units = single]{0.7;10}{\milli\tesla} is considered. The information needed to generate TF-$\mu$SR histograms is: \begin{itemize}
    \item the fraction of implanted muons that exist in a neutral state \Muz at a time $t=0$. It is assumed that \FMu($t=0$)=0.95, because at low temperatures a diamagnetic fraction of 0.05 is measured in 4H-SiC with \LEmuSR.
    \item the four precession frequencies $\nu_{ij}$ and respective probabilities $a_{ij}$ for each simulated magnetic field. The transition frequencies used, and their probabilities, are shown for \SIlist[list-units = single]{0.7;10}{\milli\tesla} in \Cref{tab:Mutransitions}.
    \item the rate of the capture process. The simulation was run for different $\lambda_c$ in the range of \SIrange[range-units = single]{0.1}{50000}{\per\micro\second}.
\end{itemize}

The time $t_c$ of the next electron capture event depends on $\lambda_c$ and is "thrown" according to the probability distribution $\exp(-\lambda_ct)$. 
In the \Muz state, the muon spin precession is determined by the superposition of the precession frequencies $\nu_{ij}$, weighted by their respective probabilities $a_{ij}$. 
If $t_c > t_d$, with $t_d$ the muon decay time, the muon decays while still being in the \Muz state. If $t_c < t_d$, the electron capture process results in the final diamagnetic state \Mum, and the muon spin continues to precess at its Larmor frequency at $t > t_c$ until it decays at $t_d$. 
More information on the Monte-Carlo simulation methodology can be found in Refs.~\cite{Prokscha_2012,Prokscha_2014}.

\begin{table}[t]
\caption{\label{tab:Mutransitions} Transition frequencies $\nu_{ij}$ and probabilities $a_{ij}$ of \Muz in 4H-SiC in an applied TF of \SI{0.7}{\milli\tesla} and \SI{10}{\milli\tesla}, assuming an isotropic hyperfine coupling constant A$_\textrm{HF}$ = \SI{3030}{MHz}, corresponding to the \Muone state in 4H-SiC.}
\begin{ruledtabular}
    \begin{tabular}{ccccc}
    \multirow{2}{*}{Transition} & \multicolumn{2}{c}{0.7 mT} & \multicolumn{2}{c}{10 mT} \\
     & $\nu_{ij}$ (MHz) & $a_{ij}$ & $\nu_{ij}$ (MHz) & $a_{ij}$ \\
\hline
   $\nu_{12}$      & 9.7 & 0.252  & 132.7 & 0.273 \\
    $\nu_{23}$     & 9.8 &0.248 & 145.8 & 0.227 \\ 
    $\nu_{34}$     & 3020.2 & 0.252  & 2897.2 & 0.273 \\
    $\nu_{14}$    & 3039.7& 0.248 & 3175.7 & 0.227\\ 
    
    \end{tabular}
    \end{ruledtabular}
\end{table}

\section{\label{sec:RnD}Results and discussion}

\begin{figure*}[hbt!]

\includegraphics[width=0.95\textwidth]{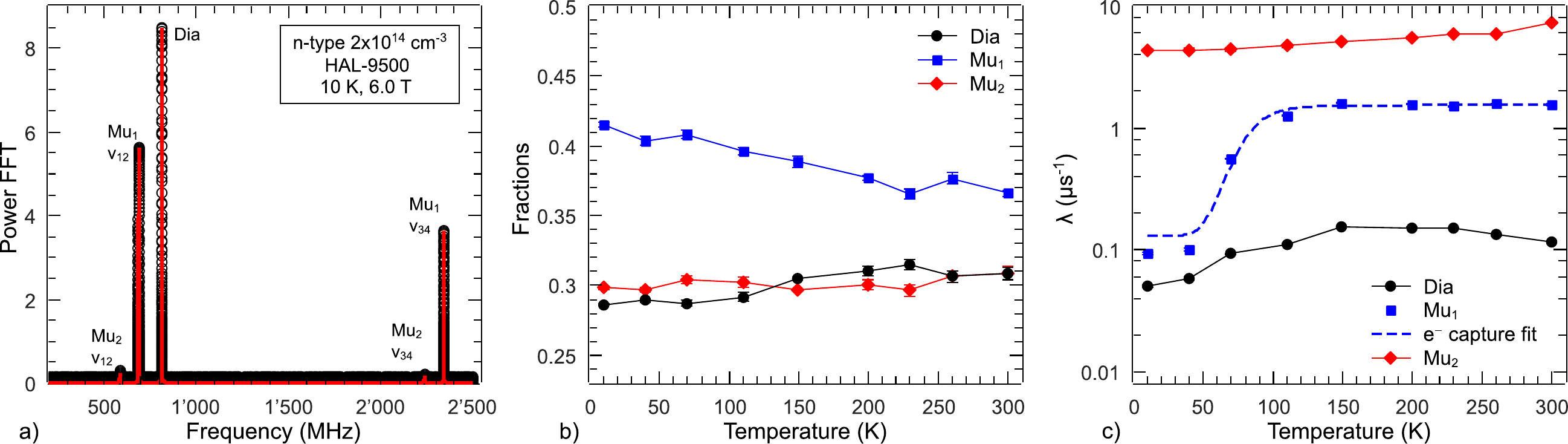}% Here is how to import EPS art
\caption{\label{fig:HAL-Tscan} a) Fourier transform power spectrum of the time spectrum collected in HAL-9500. The \Muone and \Mutwo states, with distinct hyperfine
coupling frequencies are observed, in addition to the
diamagnetic line in the bulk of thick epilayer at \SI{10}{\kelvin} and \SI{6}{\tesla}. 
Results from HAL-9500 measurements in n-type ($2\times 10^{14}$~cm$^{-3}$, N-doped) 4H-SiC with an applied magnetic field of \SI{6}{\tesla}. Temperature dependence of b) the fractions of the diamagnetic (Dia), and paramagnetic \Muone and \Mutwo components, and c) the corresponding relaxation rates $\lambda$. The solid lines in b,c) are guides to the eyes, and the dashed line is the fit curve using \Cref{eq:lambda_fit_final} to model the \Muone electron capture process. }
\end{figure*}

The high field (\SI{6}{\tesla}) measurements performed in HAL-9500 confirm the presence of the two \Muz configurations previously observed in 4H-SiC \cite{Lichti_2004}, as can be seen in \Cref{fig:HAL-Tscan}~a).

\begin{figure*}[hbt!]

\includegraphics[width=0.99\textwidth]{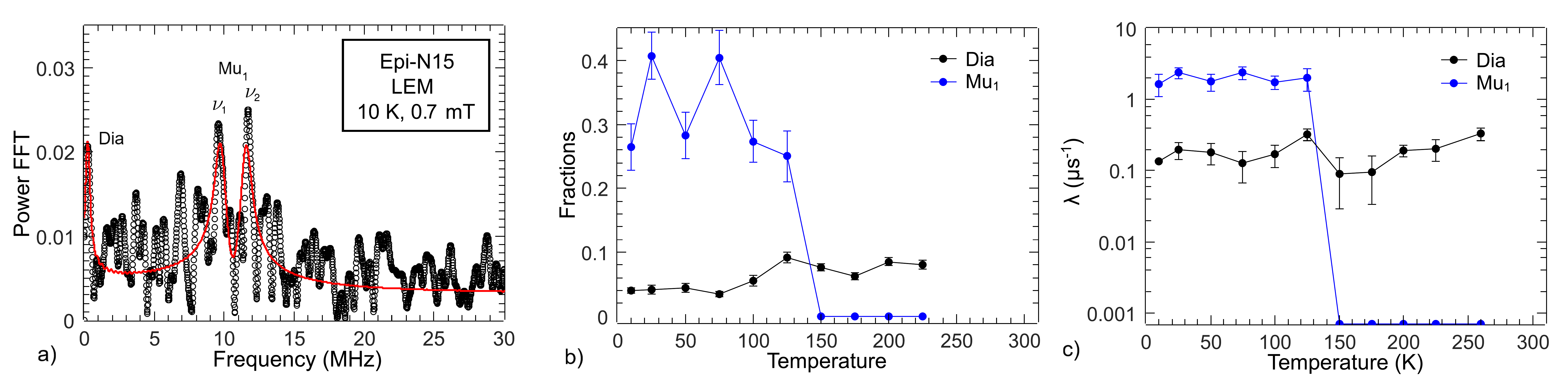}% Here is how to import EPS art
\caption{\label{fig:low-dopedSiCMu_LEM} a) Fourier transform of the time spectrum collected
in LEM. In the Epi-N15 sample the diamagnetic and \Muone precession frequencies are observed at \SI{10}{\kelvin} and \SI{0.7}{\milli\tesla}.
The $\nu_{12}$ and $\nu_{23}$ transitions of the \Muz state are split into
two lines, with frequencies $\nu_1=$~\SI{9.8}{\mega\hertz} and $\nu_2=$~\SI{11.5}{\mega\hertz},
respectively. \LEmuSR measurement in Epi-N15 of b) the diamagnetic (Dia), and paramagnetic \Muone fractions and c) the respective
relaxation rates $\lambda$ as a function of temperature, measured at
\SI{0.7}{\milli\tesla} and an implantation energy of \SI{19}{\kilo\electronvolt}. The \Muone relaxation rate becomes too high above 150~K, which leads to the disappearance of the \Muone precession lines.
For simplicity, the Mu parameters were fixed at zero in this range.}
\end{figure*}

 The \LEmuSR measurements at low magnetic field (B<\SI{1}{\milli\tesla}) allow observing the lower frequency lines of the \Muz, since the upper detection limit is about \SI{60}{\mega\hertz}. Interestingly, only one of the \Muz states is observed in SiC, as seen for sample Epi-N15 in \Cref{fig:low-dopedSiCMu_LEM}~a). Since the \Muone and \Mutwo have similar hyperfine constants, an overlap of the corresponding frequency lines at \SI{0.7}{\milli\tesla} is expected, but only the component corresponding to the \Muone is observable: the depolarization rate of
 the \Mutwo state is with $> 4$~$\mu$s$^{-1}$ (see \Cref{fig:HAL-Tscan}~c)) too large to be observable in the LE-$\mu$SR spectra with their limited statistics.

Both \Muz centers have previously been reported to be isotropic \cite{Celebi_2009}, however, the presence of two \Muone lines at \SI{0.7}{\milli\tesla} for $\nu_{12}$ and $\nu_{23}$ ($\nu_{1}=\SI{9.8}{\mega\hertz}$ and $\nu_{2}=\SI{11.5}{\mega\hertz}$) indicates a small anisotropy of $\sim 1.7$~MHz of the \Muone hyperfine coupling \cite{Martins_2023}, much smaller than the isotropic part of $\approx$~\SI{3030}{\mega\hertz} \cite{Lichti_2004}. The splitting of the muonium lines is observed for all samples in the low-field and low temperature (\SI{10}{\kelvin}) \LEmuSR measurements.


\subsection{Temperature dependence}

The temperature dependence investigated in HAL-9500 and in LEM gives insight into the thermally activated processes and charge transitions of muonium in SiC in the temperature range between \SIlist{10;300}{\kelvin}.

 The temperature dependence of the muonium formation at \SI{6}{\tesla} in the bulk of the n-type epilayer with N$_D$=\SI{2e14}{\per\cubic\centi\meter} is shown in \Cref{fig:HAL-Tscan}~b,c). At \SI{10}{\kelvin} about \SI{100}{\percent} of the muon polarization is observable, with \SI{29}{\percent} being in a diamagnetic state, \SI{42}{\percent} in \Muone and \SI{30}{\percent} in \Mutwo. At $\sim$\SI{50}{\kelvin} the thermal ionization of the N donors starts \cite{Kimoto_1995a}, leading to the increase of the relaxation rates $\lambda$ of the diamagnetic and \Muone components, and simultaneously the \Muone fraction decreases. The \Muone center has been assigned to the T$_\text{Si}$ site next to a Si atom \cite{Lichti_2004,Woerle_2020}, where both neutral and negatively charged hydrogen configurations are stable \cite{Kaukonen_2003, Bathen_2020}. Considering the acceptor character of \Muone, electron capture is expected to take place above \SI{50}{\kelvin}, without site change \cite{BaniSalameh_2007}. The variation of the \Muone relaxation rate $\lambda_\text{Mu$_1$}$ with temperature was fitted using \Cref{eq:lambda_fit_final} (dashed line in \Cref{fig:HAL-Tscan}~c))
 %(shown in \cref{fig:fit_lambdaMu1_alternative})) 
 to extract the capture cross section  $\sigma^{0/-}$ of the electron capture process. More details about the fit are included in the \Cref{sec:Appendix_ElectronCapture} .
 The decrease in diamagnetic fraction just above \SI{200}{\kelvin} (in \Cref{fig:HAL-Tscan}~b)) points to the ionization of the \Mum state with a site change to form \Mutwo \cite{Celebi_2009}, whereby the \Mutwo has been assigned to the anti-bonding site to a carbon atom \cite{Celebi_2009,BaniSalameh_2007}.

The temperature dependence of the different muonium fractions was measured with \LEmuSR at \SI{0.7}{\milli\tesla} and \SI{19}{\kilo\electronvolt}. For the sample with the lowest doping of \ND$=\SI{4e15}{\per\cubic\centi\meter}$ (Epi-N15) a small increase of the diamagnetic fraction with temperature was observed in \Cref{fig:low-dopedSiCMu_LEM}~b), between \SIlist[list-units = single]{10;260}{\kelvin}. For this electron concentration, the capture rate $\lambda_c$ ruling the conversion from \Muz to \Mum is comparable to the lower muonium frequencies, leading to a loss of the initial muon polarization \cite{percival_radiolysis_1978} seen in the disappearance of the \Muone components (in \Cref{fig:low-dopedSiCMu_LEM}~b-c)) as its precession becomes unresolved.
Thus, the conversion process is not fully observable for this low doping concentration in the SiC.
As previously observed, typical \FD values are below $0.1$ \cite{Woerle_2020}, with the remaining \textit{missing} fraction arising from the fraction of implanted \mup that undergo fast depolarization effects.
\begin{figure*}[hbt!]

\includegraphics[width=\textwidth]{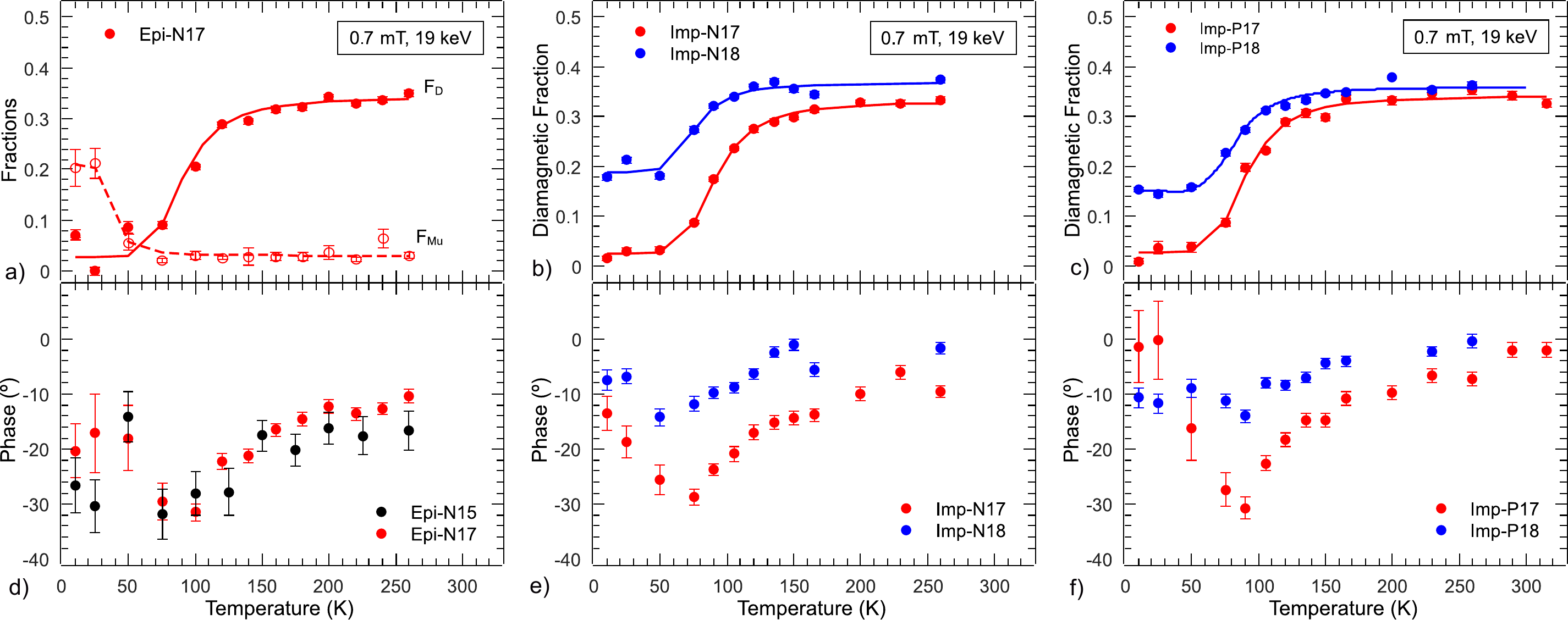} % Here is how to import EPS art
\caption{\label{fig:nSiC_Tscan} Muonium fraction \FMu, diamagnetic fraction \FD, and phase $\phi_\text{D}$ measured as a function of temperature for (a) Epi-N17.
Diamagnetic fraction \FD and phase $\phi_\text{D}$ measured as a function of temperature for (b) Imp-N17 and Imp-N18, (c) Imp-P17 and Imp-P18. The solid lines show the results of the fits using \Cref{eq:BoltzmannTmodel} to extract the activation energy of the SiC donor species.}
\end{figure*}

For the samples with higher doping, the \LEmuSR results of the temperature dependence of the diamagnetic and paramagnetic fractions (\FD and \FMu), and phase $\phi_\text{D}$ are shown in \Cref{fig:nSiC_Tscan}.
 At low temperatures in the Epi-N17 sample, \SI{20}{\percent} of muons occupy a \Muone state, and \SI{9}{\percent} form a diamagnetic state, likely \Mup. As seen with high-field \muSR, conversion of \FMu to \FD occurs above \SI{50}{\kelvin}, as the donors in SiC start to ionize, and the availability of free carriers contributes to the formation of \Mum via electron capture. This conversion is seen in \Cref{fig:nSiC_Tscan}~a) as \FMu drops to zero at \SI{50}{\kelvin}, while \FD gradually increases between \SIlist[list-units = single]{50;125}{\kelvin}, until the donor ionization is complete and \Mum formation saturates.
The formation of \Mum usually requires a neutral precursor, which in the case of 4H-SiC is the acceptor \Muone, and the process is thus said to be \textit{delayed}. A significant indication of the \textit{delayed} electron capture process is the negative shift in phase (\Cref{fig:nSiC_Tscan}~d-f)) \cite{Patterson_1988}, which is not expected when the diamagnetic state is promptly formed, for example, if the implanted \mup remains in the positive \Mup state until the decay.
The phase shift appears for all samples  (in \Cref{fig:nSiC_Tscan}~d-f)) but recovers faster (i.e. is less negative) for the highly doped samples Imp-N18 and Imp-P18. This follows the description of \Cref{eq:lambda_fit} that electron capture becomes faster with increasing electron availability.

The thermally activated increase of \FD, observed for all samples except Epi-N15, can be fitted with the Boltzmann model from \Cref{eq:BoltzmannTmodel} \cite{Vilao_2017,Alberto_2018a}, to extract the activation energy $E_\text{A}$ of the charge transition process.
\begin{equation}
\label{eq:BoltzmannTmodel}
    \text{F}_\text{D} (\text{T}) =f_0+\frac{b_0 \cdot N \cdot e^{-\frac{E_\text{A}}{k_B T}}}{1+N\cdot e^{-\frac{E_\text{A}}{k_B T}}}
\end{equation}
where  $f_0$ is the diamagnetic fraction at low temperature ($<$ \SI{50}{\kelvin}), $N$ is a density-of-states parameter, $b_0$ is the interconverting fraction, and $k_B$ is the Boltzmann constant. The values of $N$ used in the fits were fixed to $N=558$ and $N=581$ for the epilayer and ion implanted samples, respectively. These $N$ values are the average of the %previously obtained 
values from fits with free parameters. This model assumes that the temperature provides the activation energy, which is reasonable in the case of the \Muz to \Mum transition because it requires the presence of free electrons, made available by the thermal ionization of the donors in 4H-SiC. Thus, it is expected that the extracted \EA agrees with the energy level of N or P donors in 4H-SiC. The ionization energy depends on whether the donor atoms occupy a cubic (\textit{k}) or a hexagonal (\textit{h}) site. For N the energies are 61.4/126~meV and for P 120/60.7~meV in \textit{h/k} sites, respectively \cite{Ivanov_2005, Kimoto_2014Fund}.

\begin{table}[t]
\caption{\label{tab:TscanEA} Diamagnetic fractions \FD measured at 
10~K and 260~K at an implantation energy E=\SI{19}{\kilo\electronvolt} and a magnetic field B=\SI{0.7}{\milli\tesla}. The donor activation energy $E_\text{A}$ is derived by fitting \Cref{eq:BoltzmannTmodel} to \FD(T) in \Cref{fig:nSiC_Tscan}~a-c). }
\begin{ruledtabular}
    \begin{tabular}{cccc}
    \textbf{Sample name} & \textbf{$E_\text{A}$} (meV) & \textbf{ \FD (\SI{10}{\kelvin})} & \textbf{\FD (\SI{260}{\kelvin})}\\
\hline
   Epi-N15      & $-$  & \textcolor{black}{0.044(4)}  & \textcolor{black}{0.091(13)} \\
    Epi-N17     & 61(9) & \textcolor{black}{0.071(10)} & \textcolor{black}{0.350(7)} \\ \hline
    Imp-N17     & 57(2) & \textcolor{black}{0.016(5)}  & \textcolor{black}{0.334(6)} \\
    Imp-N18    & 48(5) &\textcolor{black}{0.178(6)} & \textcolor{black}{0.374(6)}\\ \hline
    Imp-P17     & 57(3) & \textcolor{black}{0.0094(5)} & \textcolor{black}{0.353(9)}\\
    Imp-P18     & 53(3) & \textcolor{black}{0.154(5)} & \textcolor{black}{0.365(8)} \\
    
    \end{tabular}
    \end{ruledtabular}
\end{table}
The \EA values obtained from the fits for the different samples are shown in \Cref{tab:TscanEA}. Only one energy could be extracted, which is assigned to the ionization of the shallower donors, for N in the hexagonal C-site and P in the cubic Si-site. 

Under the given measurement conditions, with an applied field of \SI{0.7}{\milli\tesla}, we do not observe an increase in the charge carrier concentration due to the ionization of the deeper donors (which are expected to be occupied \cite{Kimoto_2014Fund}). This is because neither \FD nor $\phi_\textrm{D}$ change significantly within the concentration range of \SI{1e17}{\per\cubic\centi\meter} -- \SI{1e18}{\per\cubic\centi\meter}. This follows from the simulation discussed below (see  \Cref{fig:Simulation} b)).
The \EA for samples with \SI{1e18}{\per\cubic\centi\meter} doping is lower than expected; however, it is in agreement with the electrical characteristics observed in P-implanted SiC, where the ionization energy level decreases with increasing doping concentration \cite{Das_2024}.

The samples Imp-N18 and Imp-P18 show an increase in \FD compared to Imp-N17 and Imp-P17 at low temperatures, which indicates that more damage is created by the higher fluence implantation (about one order of magnitude larger for Imp-18 than for Imp-17). It is presumed that \FD increases due to either permanent damage created during the implementation, not recoverable by the PIA in Imp-N18 and Imp-P18, or that more atoms are displaced which contribute to an increase of C-related defects during PIA, as seen with \LEmuSR in aluminum implanted samples \cite{Kumar_2024_AlImp}. The former explanation implies that the higher fluence implantation leads to some amorphization of the lattice, where the likelihood of the incoming muons to stop in a site suitable to form \Muz is reduced, hence increasing \FD. The latter involves the displacement of C atoms by the impinging ions, creating \VC, which are not fully annealed out even with the use of the C-cap during annealing, and \FD increases due to the formation of (\VC-Mu)$^-$ complexes \cite{Woerle_2020}.

\subsection{Doping profiles after N and P implantation}

\begin{figure}[hbt!]
\hspace{-0.3cm}
\includegraphics[width=0.4 \textwidth]{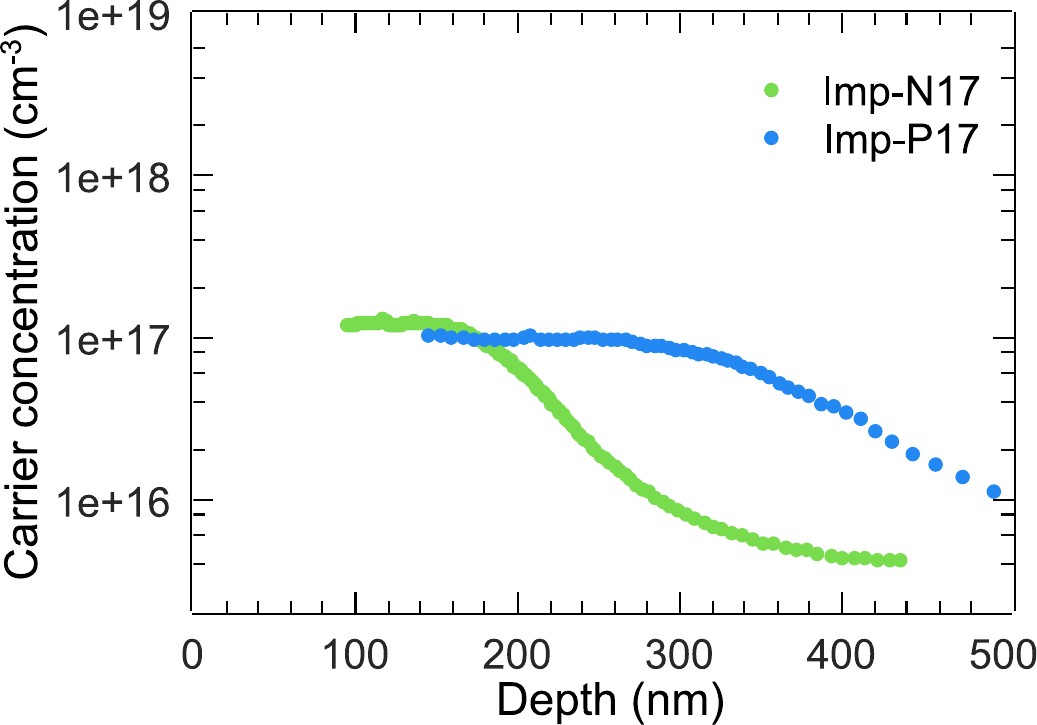}
\caption{\label{fig:CV-doping} Depth profiles of the doping concentrations in samples Imp-N17 and Imp-P17, respectively, obtained with capacitance-voltage measurement.}
\end{figure}

The depth profiles of the charge carrier concentration were obtained with capacitance-voltage (C-V) measurements, while secondary ion mass spectrometry (SIMS) was used to obtain the implanted atom density. The carrier concentration extracted from the C-V is shown in \Cref{fig:CV-doping}, however, the depth profiles are only measurable for depths higher than \SI{100}{\nano\meter}, and no information on the activation of the dopants near the surface is obtained. For Imp-N17 and Imp-P17, the depth profile of the carrier concentration matches the results of the SRIM simulation (in \Cref{fig:SRIMImp-N-P}), and reveals that the dopant activation is close to \SI{100}{\percent}. Note that the C-V measurements could not be performed on Imp-N18 and Imp-P18 due to a large leakage current; therefore, the electrical activation could not be obtained.

\begin{figure}[hbt!]
\hspace{-0.3cm}
\includegraphics[width=0.4\textwidth]{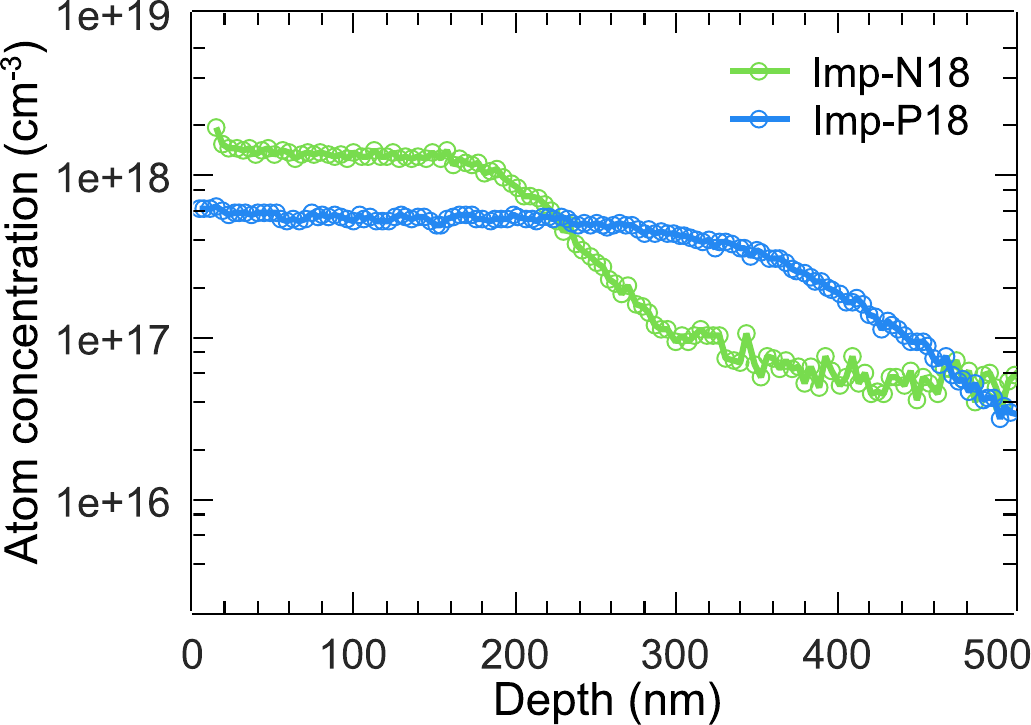}
\caption{\label{fig:SIMS} Depth profiles of N and P concentration in samples Imp-N18 and Imp-P18, respectively, obtained from SIMS. }
\end{figure}

\Cref{fig:SIMS} shows the depth profile of the concentrations of N and P atoms in Imp-N18 and Imp-P18, respectively. The depth profiles of the samples Imp-N17 and Imp-P17 could not be accurately determined because the dopant concentration was too close to the lower detection limit of the SIMS tool. Interestingly, the N concentration in the box profile is about twice the concentration of P.

\subsection{\LEmuSR depth analysis}
\begin{figure*}[hbt!]
\hspace{-0.3cm}
\includegraphics[width=\textwidth]{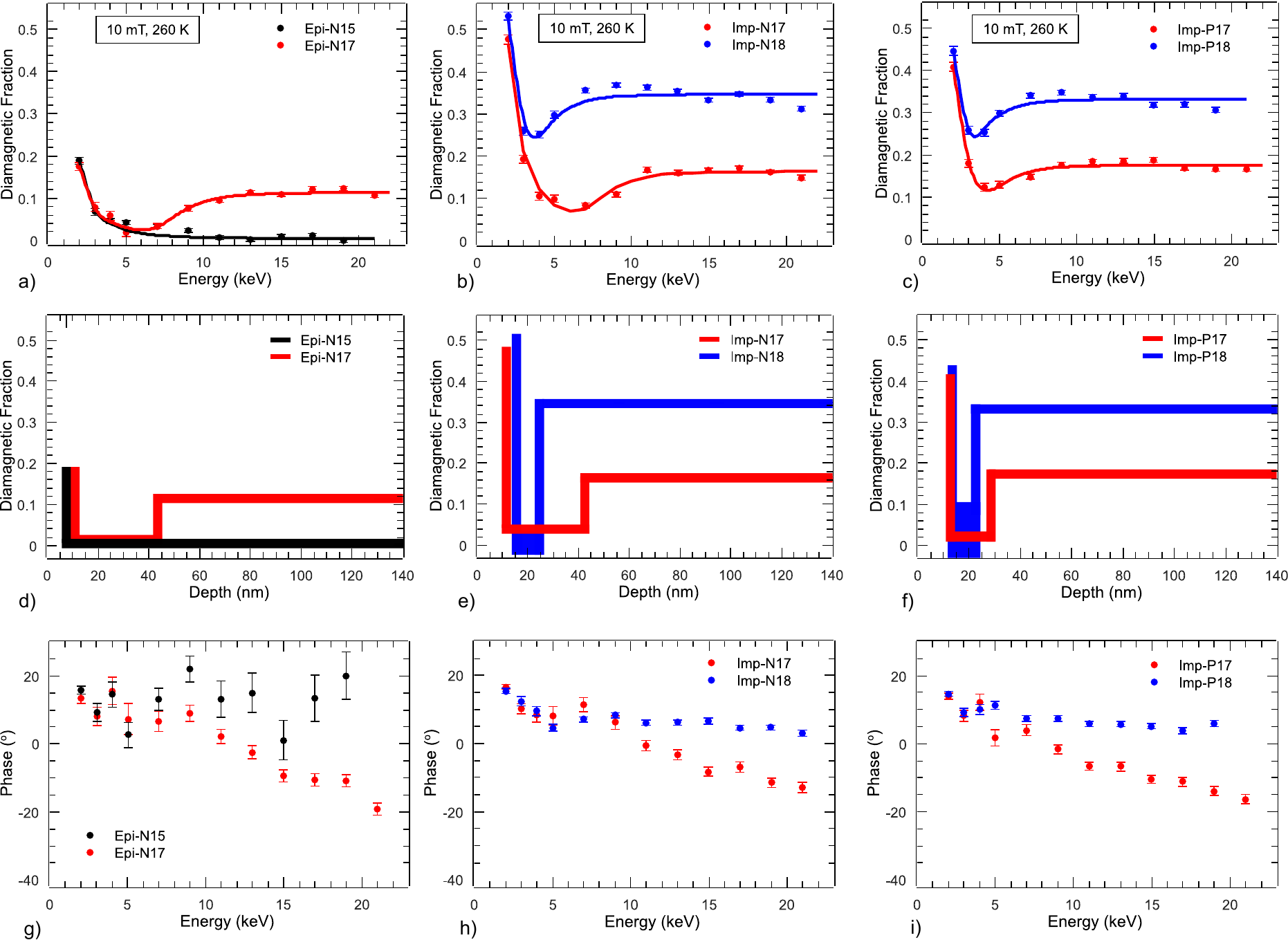}
\caption{\label{fig:nSiC_Steps} Diamagnetic fraction \FD as a function of energy and fitted \cite{Simoes_2020,martins_depth_2023} assuming the respective depth variation below for a) epilayer, b) N-implanted and c) P-implanted samples. d-f) The thickness of the lines of the depth variation models represents the standard deviation of the fit parameters. g-i) phase $\phi_\text{D}$ of the diamagnetic signal measured as a function of muon implantation depth. Note, that in an applied field of \SI{10}{\milli\tesla}, in the absence of muon charge state conversions, $\phi_\textrm{D}$ is about 15$^\circ$, compared to about 0$^\circ$ in \SI{0.7}{\milli\tesla} in \Cref{fig:nSiC_Tscan}. This difference arises from the different spin precession angles of the low-energy muons in the fringe field of the experimental magnet as they approach the sample.}
\end{figure*}

The energy dependence of the muon signal was measured at \SI{260}{\kelvin}, and the fraction and phase of the diamagnetic component are shown in \Cref{fig:nSiC_Steps}. Although the donors are completely ionized at this temperature, the variation of the \FD with depth suggests the free carrier concentration is not homogeneous over the first \SI{150}{\nano\meter}. For all samples, an increase in \FD at the sample surface is observed. 
During the slowing down process of the muon, electron-hole pairs are created \cite{Emery_1965}. However, at low muon implantation energies (E<\SI{10}{\kilo\electronvolt}) less of these track electrons are generated, which results in an increase in \FD at the sample surface due to the reduced probability to form
\Muz from electron capture, and the final state of the muon is \Mup \cite{Prokscha_2007,Woerle_2019a,Eshchenko_2002}. The increase in \FD at the surface depends on the sample, being \SI{20}{\percent} in the epilayer samples 
(\Cref{fig:nSiC_Steps}~a)) and \SIrange{40}{50}{\percent} in the implanted samples (\Cref{fig:nSiC_Steps}~b,c)). This is attributed to the damage created near the sample surface during the implantation process \cite{Martins_2023}, and additional degradation at the SiC surface due to the annealing process with the C-cap \cite{Linnarsson_2018}. Furthermore, a thin native oxide layer can form in 4H-SiC due to air exposure of the sample surface\cite{amy2002si,daSilva_2004,Huang_2012}. 
The presence of a native oxide on all samples could explain the reduction of \FD setting in at a depth of $\sim$\SI{10}{\nano\meter}. In this region, the \FD signal is not increasing with temperature, suggesting depletion of electrons. This is unexpected because no metal was deposited on the investigated structures, which would induce band-bending due to the difference in work functions of the different materials. However, band-bending in the region between \SIlist[list-units = single]{10;40}{\nano\meter} could be induced by charge capture at the SiC surface, possibly in the thin native oxide layer \cite{monch_semiconductor_surfaces_2001}. Interestingly, the width of the region with reduced \FD depends on the doping concentration of the sample, as expected in the case of electron depletion. The depletion width (\WD) values were extracted from the depth models in \Cref{fig:nSiC_Steps}~d-f), and are summarized in \Cref{tab:WD_100G}. The electron depletion was not observed for Epi-N15, since \LEmuSR is not sensitive to an electron concentration as low as \SI{4e15}{\per\cubic\centi\meter} in 4H-SiC. The \WD is expected to depend on the doping concentration, which is seen in the similar values for Epi-N17 and Imp-N17, and the samples with higher doping Imp-N18 and Imp-P18. However, the extracted \WD for Imp-P17 is considerably smaller than for the N17 samples, and is likely due to the difference in the behavior of N and P in the SiC lattice.
The presence of \VC with a density greater than \SI{1e17}{\per\cubic\centi\meter} increases \FD \cite{Woerle_2020}.  
Generation of \VC during ion implantation of Al has been investigated with \LEmuSR, revealing that \FD increases with \VC concentration \cite{Kumar_2024_AlImp}. Carbon displacement can occur also due to N- and P-implantation, however incorporation of the N-atoms should result in the occupation of the C-site, N$_\text{C}$ \cite{rauls_different_2003,erlekampf_deeper_2019}, while P-donors are incorporated as P$_\text{Si}$. Thus, the increase of \FD due to the presence of unpassivated \VC in Imp-P17 with the formation of \VC-\Mum complex does not allow for an accurate extraction of \WD for this sample, since it requires decoupling of the contributions of the C-vacancies and the free electrons to \FD.
\textcolor{black}{Furthermore, small deviations between the experimental profiles and the depletion-layer model are observed at larger implantation energies, particularly for the highest-fluence samples Imp-N18 and Imp-P18. Since these deviations become more pronounced with increasing implantation fluence and are accompanied by an enhanced \FD below the depletion region, they are attributed to implantation-induced lattice damage, likely involving the formation of \VC.} Beyond the depletion region, the muon signal depends on the doping concentration, as the increase of \FD is accompanied by a phase shift (\Cref{fig:nSiC_Steps}~g-i)) indicating delayed \Mum formation in the presence of free electrons.

\begin{table}[t]
\caption{\label{tab:WD_100G} Depletion width \WD and diamagnetic fraction \FD recorded at T=\SI{260}{\kelvin} and B=\SI{10}{\milli\tesla}. \WD is the width of the region where a drop in \FD is visible.}
\begin{ruledtabular}
    \begin{tabular}{ccc}
    \textbf{Sample name} & \WD (\SI{}{\nano\meter})  & \textbf{\FD (\SI{19}{\kilo\electronvolt})}\\
\hline
   Epi-N15      & $-$  &  \textcolor{black}{0.014(4)}
 \\
    Epi-N17     & 34(3)&  \textcolor{black}{0.123(6)} \\ \hline
    Imp-N17     & 31(1) &  \textcolor{black}{0.163(4)} \\
    Imp-N18    & 8(1) &  \textcolor{black}{0.334(5)}\\ \hline
    Imp-P17     & 16(1) &  \textcolor{black}{0.167(6)}\\
    Imp-P18     & 9(2) &  \textcolor{black}{0.319(6)} \\
    \end{tabular}
    \end{ruledtabular}
\end{table}

\subsection{Electron capture of Mu in n-type 4H-SiC}\label{sec:electron-capture}

\begin{figure}[hbt!]
%\hspace{-0.3cm}
\includegraphics[width=0.42\textwidth]{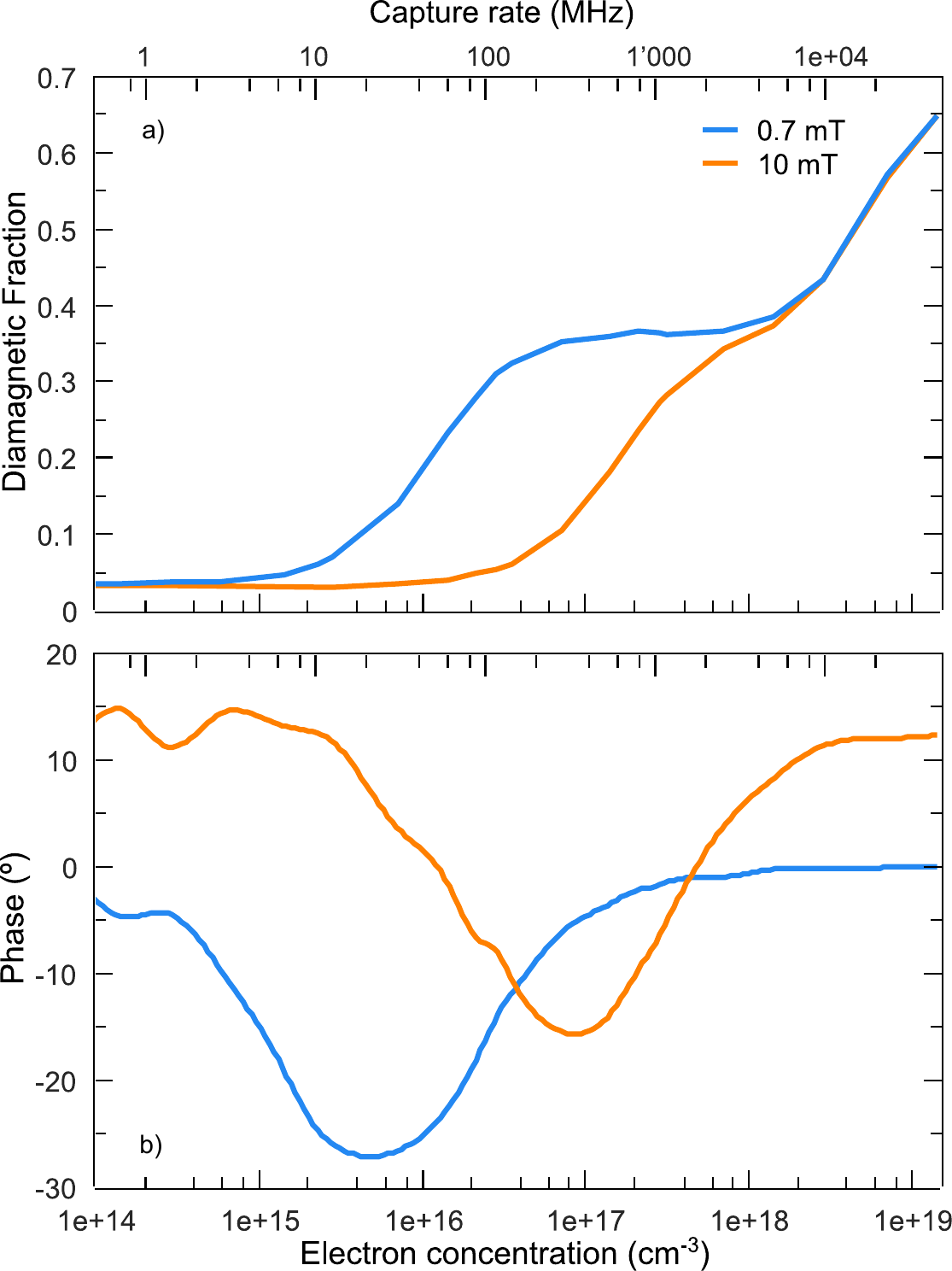}% Here is how to import EPS art
\caption{\label{fig:Simulation} Simulated dependence of the a) diamagnetic fraction \FD and b) phase $\phi_\textrm{D}$ on the free electron concentration $n$ (at \SI{260}{\kelvin}) and capture rate $\lambda_c$ for the \Muone state with an hyperfine coupling constant A$_\textrm{HF}$ = \SI{3030}{MHz}. At \SI{10}{\milli\tesla}, we added in the
simulation an offset of $\phi_\text{D} = 15^\circ$, to account for the observed spin precession of the muons in the fringe field of the experimental magnet on approaching the sample.}
\end{figure}

 With increasing doping concentration, and thus effective free electron concentration, the electron capture process $\text{Mu$^0$} + \text{e}^- \rightarrow \text{Mu}^-$ leads to an increase of \FD and a negative shift of $\phi_\textrm{D}$ at \SI{260}{\kelvin}. To model the electron capture process, we employed the simple Monte-Carlo simulation described in \Cref{subsec:simulation} to calculate $\mu$SR time spectra with statistics similar to the experiment and determined \FD and $\phi_\textrm{D}$ by fitting the simulated data. 
In the simulation, only the state transitions involving the \Muone state are considered, since it behaves as an acceptor, and as seen in \Cref{fig:HAL-Tscan}, the electron capture by \Muone is the main contributor to the increase of the diamagnetic fraction in n-type SiC.

The diamagnetic fraction \FD and phase $\phi_\textrm{D}$ measured with \LEmuSR for different doping concentrations (from \Cref{tab:Samples}) were used to calibrate the simulated curves shown in \Cref{fig:Simulation}. The diamagnetic fraction (\Cref{fig:Simulation}~a)) and phase (\Cref{fig:Simulation}~b)) are extracted as a function of capture rate $\lambda_c$, and the electron concentration is then calculated from \Cref{eq:lambda_fit}. To match the collected \FD and $\phi_\textrm{D}$ to the known free electron concentration, the relation $\sigma^{0/-} \cdot v_e= \SI{3.5e-9}{\cubic\centi\meter\per\second}$ was found at a temperature of \SI{260}{\kelvin}.

At \SI{260}{\kelvin}, the thermal velocity $v_e$ of electrons for an effective electron mass of $0.42\cdot m_e$ is $\sim 1.7\times10^7$~cm/s (\Cref{sec:Appendix_ElectronCapture}). This allows us to estimate the electron capture cross section at \SI{260}{\kelvin} to be $\sigma^{0/-}\sim\SI{2e-16}{\centi\meter\squared}$, in excellent agreement with the
procedure described in \Cref{sec:Appendix_ElectronCapture}.

Interestingly, the \SI{10}{\milli\tesla} simulated curves of \FD and $\phi_\textrm{D}$ are shifted to the right compared to the \SI{0.7}{\milli\tesla} data, and the increase of \FD starts only at $n\gtrsim\SI{4e16}{\per\cubic\centi\meter}$, while the presence of $n\gtrsim\SI{2e15}{\per\cubic\centi\meter}$ already induces a measurable \Mum fraction at \SI{0.7}{\milli\tesla}. This is because a capture rate slower than the \Muz frequencies results in the loss of part of the initial muon spin polarization \cite{percival_radiolysis_1978}.
The low frequency precession lines at \SI{0.7}{\milli\tesla} allow to detect the slower electron capture with $\lambda_c$ in the order of \SI{10}{\mega\hertz}, while $\lambda_c \sim$~\SI{200}{\mega\hertz} is in the range of the $\nu_{12}$ and $\nu_{23}$ frequencies at \SI{10}{\milli\tesla}. When electron capture becomes even faster, for $n>$\SI{1e18}{\per\cubic\centi\meter}, 
%both magnetic fields can detect the reaction via the high transition frequencies $\nu_{34}$ and $\nu_{14}$ of $\sim$\SI{3000}{\mega\hertz}.
we observe in both magnetic fields a simultaneous increase of \FD, where $\lambda_c$ becomes larger than the high transition frequencies $\nu_{34}$ and $\nu_{14}$ of $\sim$~\SI{3000}{\mega\hertz}.
Thus, when $\lambda_c$ is in the order of or faster than any transition frequency of \Muz (see \Cref{tab:Mutransitions})
%$\nu_{ij}$ $0.5\cdot \nu_\text{Mu} <\lambda_c < 10\cdot \nu_\text{Mu}$   
the spin coherence of the muon ensemble
is conserved, seen as an increase of \FD and a negative shift of $\phi_\text{D}$. Once the electron capture becomes sufficiently fast, the polarization loss and dephasing of the diamagnetic component are reduced, and the shift of $\phi_\text{D}$ returns to zero.

The simulated curves of the diamagnetic fraction \FD and phase $\phi_\text{D}$ can be used \textcolor{black}{as calibration curves}
to quantify the presence of free charge carriers.  \textcolor{black}{After establishing the relationship between electron capture rate and carrier concentration using reference samples with known charge carrier density, measurements on an unknown sample can be compared with simulated \FD and phase values obtained under the same magnetic field conditions.}
Using the results for both \SIlist[list-units=single]{0.7;10}{\milli\tesla} allows a reliable extraction of the effective electron concentration in the range of \SIrange[range-units=single]{2e15}{1e19} {\per\cubic\centi\meter}\textcolor{black}{, therefore providing a tool for carrier-concentration characterization in 4H-SiC.}

\section{Conclusion}

In this work, the dynamics of muonium in n-type 4H-SiC and the sensitivity of \LEmuSR to different doping concentrations in the range of \SIrange[range-units=single]{4e15}{1e18}{\per\cubic\centi\meter} were investigated. The temperature-dependent measurements allow the observation of the charge state transition \Muz to \Mum, due to electron capture. Importantly, the activation energy of this process was extracted and could be matched to the ionization levels of the donors, revealing the sensitivity of the \LEmuSR parameters to the appearance of free electron carriers. The depth-resolved analysis further uncovered electron depletion regions, likely induced by fixed charges in a native oxide layer formed at the surface of all the samples.

The samples implanted with N- and P-ions show a comparable electrical activation of the donors, seemingly \SI{100}{\percent}. However, for a doping concentration of \SI{2e17}{\per\cubic\centi\meter}, the P-implantation leads to a higher concentration of \VC, compared to N-implantation. The higher fluence used for the doping concentration \SI{1e18}{\per\cubic\centi\meter} also results in permanent damage in the lattice, not recoverable by PIA for both N- and P-implantation, with the main defect candidates being \VC created during ion implantation, and C-interstitials related centers during PIA.

The wide range of doping concentrations investigated in n-type 4H-SiC samples, combined with the complete electrical activation of donors following ion implantation, enabled the development of a model linking \LEmuSR observables to free electron concentration. Using a Monte-Carlo simulation to describe the electron capture process of \Muz in SiC, the variation of the diamagnetic fraction and phase was modeled as a function of electron capture rate. Since the electron capture rate depends on the free electron density, these results establish a method to extract doping levels and carrier concentrations in 4H-SiC through \LEmuSR measurements of the diamagnetic fraction and phase.

\section{Acknowledgments} 
The muon experiments were performed at the $\mu$E4/LEM beamline \cite{Prokscha_2008} and the HAL-9500 instrument of the Swiss Muon Source S$\mu$S, Paul Scherrer Institute, Villigen, Switzerland. This work was supported by the Swiss National Science Foundation under Grant No.$192218$. The work of MEB was supported by an ETH Z\"urich Postdoctoral Fellowship.

%\section{Appendix}\label{sec:Appendix}
\appendix
\section{Electron capture fit of Mu$_1$ relaxation rate}\label{sec:Appendix_ElectronCapture}

The relaxation rate $\lambda$ of the \Muone state, measured as a function of temperature $T$ in HAL-9500 (\Cref{fig:HAL-Tscan}~c), was fitted with \Cref{eq:lambda_fit}. The goal is to extract the activation energy $E_\text{A}$ of the process leading to an increase of $\lambda$ as a function of temperature, and the temperature dependent capture cross section $\sigma^{0/-}$ of the electron capture process for the reaction $\text{Mu$^0$} + \text{e}^- \rightarrow \text{Mu$^-$}$. The fit to the data was performed in the temperature range up to \SI{300}{\kelvin}.
%, as shown in \Cref{fig:fit_lambdaMu1_alternative}. 
Above \SI{200}{\kelvin}, the conversion process seems to saturate, and $\lambda$(T) is approximately constant.

%\begin{figure}[hbt!]
%\includegraphics[width=0.45\textwidth]{Figures/Rate_Mu1_eCapture_fit.pdf}
%\caption{\label{fig:fit_lambdaMu1_alternative} Relaxation rate $\lambda$ of the precession signal of \Muone measured as a function of temperature in HAL-9500 and fitted using \cref{eq:lambda_fit_final} to model the \elec capture process.}
%\end{figure}

The temperature dependence of the carrier concentration $n(T)$ depends on the ionization of the nitrogen donors with concentration $N_D = 2\times 10^{14}$~cm$^{-3}$ as a function of temperature and is determined by the following function \cite{Kimoto_2014Fund}:
\begin{equation}
    n(T) = N_D^+(T) = \frac{\gamma(T)}{2} \left( \sqrt{1 + \frac{4N_D}{\gamma(T)}} - 1 \right),
    \label{eq:nCalculation}
\end{equation}
where
\begin{equation}
    \gamma(T) = \frac{N_C(T)}{g_D} \exp\left( -\frac{E_C - E_D}{k_BT} \right),
\end{equation}
with $E_D$ the donor energy level beneath the conduction band level $E_C$, $g_D$ the degeneracy factor for donors (taken as 2), and $N_C$ the effective density of states in the conduction band given by
\begin{equation}
    N_C = 2 M_c \left( \frac{2\pi m^*_\text{e} k_BT}{h^2} \right)^{3/2}.
\end{equation}
Here, $M_c = 3$ is the number of conduction band minima in the first Brillouin zone \cite{Kimoto_2014Fund} and $h$ is Planck's constant. $n(T)$ saturates at $T>150$~K for an activation energy $E_A = E_C -E_D$ of \SI{60}{\milli\electronvolt}.

For the electron velocity $v_e$, we assume the electron thermal velocity:
\begin{equation}
     v_e(T)=\sqrt{\frac{3 k_B T}{m^*_e}},
     \label{eq:veT}
 \end{equation}
where $k_B$ is the Boltzmann constant and $m^*_e = 0.42\cdot m_e$ \cite{Kimoto_2014Fund} is the density-of-states effective mass for electrons.

Since $n(T)$ saturates at $T>$~\SI{150}{\kelvin} while $v_e(T)$ is increasing with $\sqrt{T}$, the electron capture cross section $\sigma^{0/-}$ obviously has to decrease with $1/\sqrt{T}$ in this model to yield a constant $\lambda(T)$ at $T>$~\SI{150}{\kelvin}. The \Muone data in \Cref{fig:HAL-Tscan}~c) was fitted using the function
\begin{equation}
    \lambda(T) = \lambda_c(T) = n(T)\cdot v_e(T) \cdot \sigma^{0/-}(100~\text{K})\cdot 1/\sqrt{T}.
    \label{eq:lambda_fit_final}
\end{equation}

The free parameters in \Cref{eq:lambda_fit_final} were $E_A$ and $\sigma^{0/-}(100~\text{K})$, with $E_A = $~\SI{66(5)}{\milli\electronvolt}, and $\sigma^{0/-}(100~\text{K}) = 3.4(3)\times 10^{-16}$~cm$²$.
$E_A$ fits well to the known activation energy of \SI{61.4}{\milli\electronvolt} of N at the $h$-site \cite{Kimoto_2014Fund}.

With the assumed square-root temperature dependence of $\sigma^{0/-}$, we can estimate the expected capture cross section at \SI{260}{\kelvin} to be $\sigma^{0/-}(260~\text{K}) = \sqrt{100/260}\cdot \sigma^{0/-}(100~\text{K}) \simeq 2.1\times 10^{-16}$~cm$^2$. 
%This is about two times larger than the value obtained from the estimate in \Cref{sec:electron-capture}, which was for a sample with three orders of magnitue higher doping level. This could indicate that the electron capture cross section of \Muz depends on the doping level – i.e.~disorder – in the system.
%
% Thomas: Robert suggested to add this to the appendix, which I find a good idea
%         as we have all the data already available.
%
%

\section{Temperature dependence of the hyperfine coupling of Mu in 4H-SiC}

\begin{figure}[hbt!]
\includegraphics[width=0.45\textwidth]{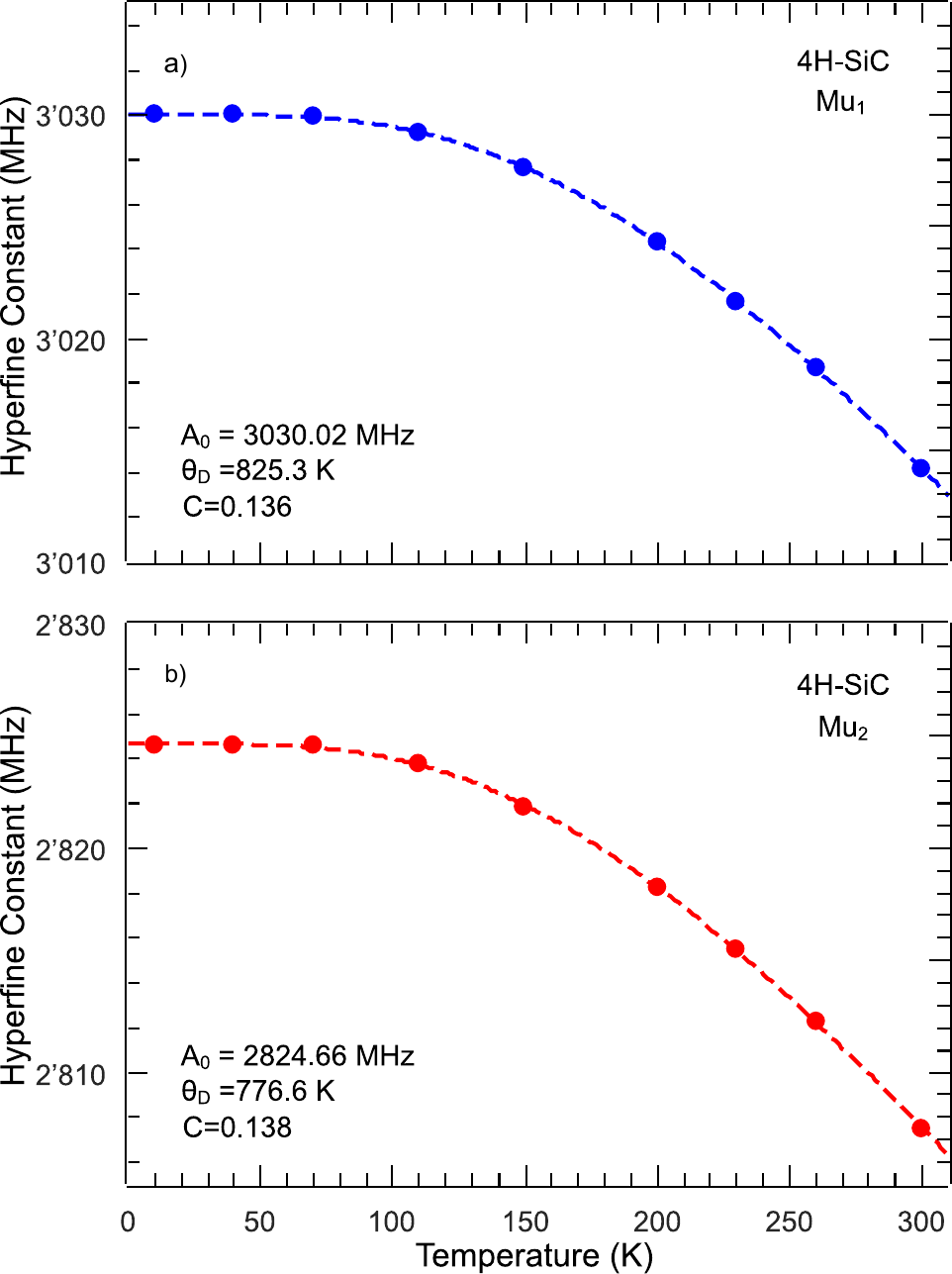}
\caption{\label{fig:debye-model} Hyperfine coupling $A_{hf}(T)$ as a function of temperature of the a) \Muone and b) \Mutwo states in n-type (\ND$=2\times 10^{14}$~cm$^{-3}$) epitaxial 4H-SiC. The dashed lines are fits of \Cref{eq:debye-model} to the data.}
\end{figure}

Although this is less relevant to the main message of this article, we would like to include the temperature dependence of the hyperfine coupling of the two muonium states observed in the HAL-9500 data to compare with previous measurements \cite{Lichti_2004}.
Data in \cite{Lichti_2004} were for bulk 4H-SiC samples with unknown doping level, indicated as high-resistive, n-type and p-type, whereas here we have a well-defined epitaxial 4H-SiC layer with n-type doping of $2\times 10^{14}$~cm$^{-3}$.

The hyperfine coupling $A_{hf}$ in \Muz is given by \cite{Patterson_1988,Blundell_2021_muon_spectroscopy,Amato_introduction_2024}
\begin{equation}
    A_{hf} = \nu_{12} + \nu_{34}, 
\end{equation}
where $\nu_{12,34}$ are the precession frequencies observed in \Cref{fig:HAL-Tscan}. As the temperature increases, the long-wavelength acoustic phonons cause a change of the overlap of the \Muz $1s$ wave function with the nearest host atoms, resulting in a reduction of $A_{hf}$ which can be calculated within the Debye model by \cite{Lichti_2004}

\begin{equation}\label{eq:debye-model}
    A_{hf}(T) = A_{hf}(0)\cdot\left[1- C\left(\frac{T}{\Theta_\text{D}}\right)^4\cdot
              \int_0^{T/\Theta_\text{D}}\frac{x^3}{e^x-1}dx\right],
\end{equation}
where $C$ is a dimensionless constant, and $\Theta_\text{D}$ is the Debye temperature. The results for the two Mu states in 4H-SiC are shown in \Cref{fig:debye-model}.
Additionally, in \cite{Lichti_2004}, the fitting of the data for n-type was not possible due to broadening of the \Muz signals above \SI{100}{\kelvin} - whereas fitting the HAL-9500 data collected between 10~K and 300~K allows extracting reliable parameters.

\begin{table}[hbt!]
\setlength{\tabcolsep}{8.5pt} 
\renewcommand{\arraystretch}{1.3} 
\caption{\label{tab:HyperfineConstant_FitParameters} Fitted parameter values for each \Muz signal in n-type 4H-SiC, using the Debye model description from \Cref{eq:debye-model}.}
\begin{ruledtabular}
    \begin{tabular}{cccc}
    Signal & A$_{hf}$(T=0) (MHz)  & $\Theta_\text{D}$ (K)  & C\\ \hline
   \Muone     & $3030.02(1)$  &   $825(9)$ & $0.136(4)$ \\
    \Mutwo     & $2824.66(3)$&  $776(18)$ & $0.138(7)$\\ 
    \end{tabular}
    \end{ruledtabular}
\end{table}

The parameters resulting from fits to the \Muz signals using \Cref{eq:debye-model} for n-type 4H-SiC are shown in \Cref{tab:HyperfineConstant_FitParameters}. These parameters are in good agreement with \cite{Lichti_2004}. A lower effective $\Theta_\text{D}$ than the literature value (from specific heat measurements) of \SI{1300}{\kelvin} \cite{Lichti_2004,Goldberg2001SiC} is found, in agreement with \cite{Lichti_2004} and studies on other semiconductors (Si, Ge) \cite{Patterson_1988}.

\noindent

% \nocite{*}
\normalem
\bibliographystyle{apsrev4-1}
\bibliography{library}

@Article{Prokscha_2007,
  title={Formation of hydrogen impurity states in silicon and insulators at low implantation energies},
  author={Prokscha, T and Morenzoni, E and Eshchenko, DG and Garifianov, N and Gl{\"u}ckler, H and Khasanov, R and Luetkens, H and Suter, A},
  journal={Physical review letters},
  volume={98},
  number={22},
  pages={227401},
  year={2007},
  publisher={APS}
}

@article{amy2002si,
  title={Si-rich 6H- and 4H-SiC (0001) 3$\times$3 surface oxidation and initial SiO$_2$/SiC interface formation from 25 to 650\SI{}{\celsius}},
  author={Amy, F. and Soukiassian, P. and Hwu, Y. K. and Brylinski, C.},
  journal={Physical Review B},
  volume={65},
  number={16},
  pages={165323},
  year={2002},
  publisher={APS}
}

@book{monch_semiconductor_surfaces_2001,
	location = {Berlin, Heidelberg},
	title = {Semiconductor Surfaces and Interfaces},
	volume = {26},
	rights = {http://www.springer.com/tdm},
	isbn = {978-3-642-08748-6 978-3-662-04459-9},
	series = {Springer Series in Surface Sciences},
	publisher = {Springer},
	author = {Mönch, Winfried},
	editorb = {Ertl, Gerhard and Gomer, Robert and Lüth, Hans and Mills, Douglas L.},
	editorbtype = {redactor},	date = {2001},
	doi = {10.1007/978-3-662-04459-9},
}

@article{kimoto_nitrogen_1995,
	title = {Nitrogen donors and deep levels in high‐quality 4H–{SiC} epilayers grown by chemical vapor deposition},
	volume = {67},
	issn = {0003-6951},
	doi = {10.1063/1.114800},
	pages = {2833--2835},
	number = {19},
	journaltitle = {Applied Physics Letters},
	shortjournal = {Applied Physics Letters},
	author = {Kimoto, T. and Itoh, A. and Matsunami, H. and Sridhara, S. and Clemen, L. L. and Devaty, R. P. and Choyke, W. J. and Dalibor, T. and Peppermüller, C. and Pensl, G.},
	date = {1995-11-06},
}

@article{percival_radiolysis_1978,
	title={Radiolysis effects in muonium chemistry},
  author={Percival, Paul W and Roduner, Emil and Fischer, Hanns},
  journal={Chemical Physics},
  volume={32},
  number={3},
  pages={353--367},
  year={1978},
  publisher={Elsevier}
}

@article{rauls_different_2003,
	title = {The different behavior of nitrogen and phosphorus as n-type dopants in {SiC}},
	volume = {340},
	issn = {0921-4526},
	url = {https://www.sciencedirect.com/science/article/pii/S0921452603006677},
	doi = {10.1016/j.physb.2003.09.234},
	series = {Proceedings of the 22nd International Conference on Defects in Semiconductors},
	pages = {184--189},
	journal = {Physica B: Condensed Matter},	author = {Rauls, E. and Gerstmann, U. and Frauenheim, Th. and Overhof, H.},	date = {2003-12-31},
	
}

@article{erlekampf_deeper_2019,
	title = {Deeper insight into lifetime-engineering in 4H-{SiC} by ion implantation},
	volume = {126},
	issn = {0021-8979},
	url = {https://doi.org/10.1063/1.5092429},
	doi = {10.1063/1.5092429},
	pages = {045701},
	number = {4},
	journal = {Journal of Applied Physics},	author = {Erlekampf, J. and Kallinger, B. and Weiße, J. and Rommel, M. and Berwian, P. and Friedrich, J. and Erlbacher, T.},	date = {2019-07-22},
}

@Book{Kimoto_2014Fund,
  title     = {Fundamentals of {S}ilicon {C}arbide {T}echnology},
  publisher = {John Wiley {\&} Sons Singapore Pte. Ltd},
  year      = {2014},
  author    = {Tsunenobu Kimoto and James A. Cooper},
  month     = {sep},
  doi       = {10.1002/9781118313534},
  url       = {https://onlinelibrary.wiley.com/doi/book/10.1002/9781118313534},
}

@INPROCEEDINGS{Kimoto_2014,
  author={Kimoto, Tsunenobu and Kawahara, Koutaro and Niwa, Hiroki and Kaji, Naoki and Suda, Jun},
  booktitle={2014 International Workshop on Junction Technology (IWJT)}, 
  title={Ion implantation technology in SiC for power device applications}, 
  year={2014},
  volume={},
  number={},
  pages={1-6},
  doi={https://doi.org/10.1109/IWJT.2014.6842018}}

@Article{Negoro_2004,
  author    = {Y. Negoro and K. Katsumoto and T. Kimoto and H. Matsunami},
  title     = {{E}lectronic behaviors of high-dose phosphorus-ion implanted {4H}-{SiC} (0001)},
  journal   = {Journal of Applied Physics},
  year      = {2004},
  volume    = {96},
  number    = {1},
  pages     = {224--228},
  month     = jul,
  doi       = {10.1063/1.1756213},
  publisher = {{AIP} Publishing},
}

@Article{Lichti_2004,
  author    = {R. L. Lichti and W. A. Nussbaum and K. H. Chow},
  title     = {Hyperfine spectroscopy of muonium in {4H} and {6H} silicon carbide},
  journal   = {Physical Review B},
  year      = {2004},
  volume    = {70},
  number    = {16},
  pages     = {165204},
  month     = oct,
  doi       = {10.1103/physrevb.70.165204},
  publisher = {American Physical Society ({APS})},
}

@Article{Eshchenko_2002,
  title={Excess electron transport and delayed muonium formation in condensed rare gases},
  author={Eshchenko, DG and Storchak, VG and Brewer, JH and Morris, GD and Cottrell, SP and Cox, SFJ},
  journal={Physical Review B},
  volume={66},
  number={3},
  pages={035105},
  year={2002},
  publisher={APS}
}

@article{Kumar_2023,
  title = {Investigation of the ${\mathrm{Si}\mathrm{O}}_{2}$-$\mathrm{Si}\mathrm{C}$ Interface Using Low-Energy Muon-Spin-Rotation Spectroscopy},
  author = {Kumar, Piyush and Martins, Maria In\^es Mendes and Bathen, Marianne Etzelm\"uller and Woerle, Judith and Prokscha, Thomas and Grossner, Ulrike},
  journal = {Phys. Rev. Appl.},
  volume = {19},
  issue = {5},
  pages = {054025},
  numpages = {17},
  year = {2023},
  month = {May},
  publisher = {American Physical Society},
  doi = {10.1103/PhysRevApplied.19.054025},
  url = {https://link.aps.org/doi/10.1103/PhysRevApplied.19.054025}
}

@article{Kumar_2024_AlImp,
title = {Al-implantation induced damage in 4H-SiC},
journal = {Materials Science in Semiconductor Processing},
volume = {174},
pages = {108241},
year = {2024},
issn = {1369-8001},
doi = {https://doi.org/10.1016/j.mssp.2024.108241},
url = {https://www.sciencedirect.com/science/article/pii/S1369800124001379},
author = {P. Kumar and M.I.M. Martins and M.E. Bathen and T. Prokscha and U. Grossner}
}

@article{Martins_2023,
author = {Mendes Martins, Maria and Kumar, Piyush and Woerle, Judith and Ni, Xiaojie and Grossner, Ulrike and Prokscha, Thomas},
title = {Defect Profiling of Oxide-Semiconductor Interfaces Using Low-Energy Muons},
journal = {Advanced Materials Interfaces},
volume = {10},
number = {21},
pages = {2300209},
doi = {https://doi.org/10.1002/admi.202300209},
year = {2023}
}

@Article{Patterson_1988,
  author    = {Bruce D. Patterson},
  title     = {Muonium states in semiconductors},
  journal   = {Reviews of Modern Physics},
  year      = {1988},
  volume    = {60},
  number    = {1},
  pages     = {69--159},
  month     = jan,
  doi       = {10.1103/revmodphys.60.69},
  publisher = {American Physical Society ({APS})}
}

@Article{Blundell_1999,
  author  = {S. J. Blundell},
  title   = {Spin-polarized muons in condensed matter physics},
  journal = {Contemporary Physics},
  year    = {1999},
  volume  = {40},
  number  = {3},
  pages   = {175--192},
  month   = may,
  doi     = {10.1080/001075199181521},
}

@book{Amato_introduction_2024,
	address = {Cham},
	series = {Lecture {Notes} in {Physics}},
	title = {Introduction to {Muon} {Spin} {Spectroscopy}: {Applications} to Solid State and {Material} Sciences},
	volume = {961},
	isbn = {978-3-031-44958-1 978-3-031-44959-8},
	shorttitle = {Introduction to {Muon} {Spin} {Spectroscopy}},
	url = {https://link.springer.com/10.1007/978-3-031-44959-8},
	urldate = {2024-03-04},
	publisher = {Springer International Publishing},
	author = {Amato, Alex and Morenzoni, Elvezio},
	year = {2024},
	doi = {10.1007/978-3-031-44959-8},
}

@Article{Bakule_2004,
  author    = {Pavel Bakule and Elvezio Morenzoni},
  title     = {Generation and applications of slow polarized muons},
  journal   = {Contemporary Physics},
  year      = {2004},
  volume    = {45},
  number    = {3},
  pages     = {203--225},
  doi       = {10.1080/00107510410001676803},
  url       = {https://doi.org/10.1080/00107510410001676803},
}

@Article{Prokscha_2008,
  author  = {T. Prokscha and E. Morenzoni and K. Deiters and F. Foroughi and D. George and R. Kobler and A. Suter and V. Vrankovic},
  title   = {The new {muE4} beam at {PSI}: {A} hybrid-type large acceptance channel for the generation of a high intensity surface-muon beam},
  journal = {Nuclear Instruments and Methods in Physics Research Section A: Accelerators, Spectrometers, Detectors and Associated Equipment},
  year    = {2008},
  volume  = {595},
  number  = {2},
  pages   = {317--331},
  month   = oct,
  doi     = {10.1016/j.nima.2008.07.081},
}

@Article{Prokscha_2014,
  author    = {Prokscha, T.},
  title     = {Simulation of {TF-muSR} and histograms in germanium in the presence of cyclic charge state transitions of muonium},
  journal   = {Journal of Physics: Conference Series},
  year      = {2014},
  volume    = {551},
  pages     = {012049},
  booktitle = {Journal of Physics: Conference Series},
}

@Article{Celebi_2009,
  author    = {Y. G. Celebi and R. L. Lichti and B. R. Carroll and P. J. C. King and S. F. J. Cox},
  title     = {Muonium in {4H} silicon carbide},
  journal   = {Physica B: Condensed Matter},
  year      = {2009},
  volume    = {404},
  number    = {23-24},
  pages     = {5117--5120},
  month     = {dec},
  doi       = {10.1016/j.physb.2009.08.246},
  publisher = {Elsevier {BV}},
}

@Article{Morenzoni_2002,
  title={Implantation studies of keV positive muons in thin metallic layers},
  author={Morenzoni, E and Gl{\"u}ckler, H and Prokscha, T and Khasanov, R and Luetkens, H and Birke, M and Forgan, E M and Niedermayer, Ch and Pleines, M},
  journal={Nuclear Instruments and Methods in Physics Research Section B: Beam Interactions with Materials and Atoms},
  volume={192},
  number={3},
  pages={254--266},
  year={2002},
  publisher={Elsevier}
}

@Book{Eckstein_1991,
  title     = {Computer {S}imulation of {I}on-{S}olid {I}nteractions},
  publisher = {Springer Berlin Heidelberg},
  year      = {1991},
  author    = {Wolfgang Eckstein},
  series    = {Springer Series in Materials Science},
  doi       = {10.1007/978-3-642-73513-4},
}

@article{Suter_2012,
  title={Musrfit: a free platform-independent framework for $\mu$SR data analysis},
  author={Suter, A. and Wojek, B. M.},
  journal={Physics Procedia},
  volume={30},
  pages={69--73},
  year={2012},
  publisher={Elsevier}
}

@article{Ayedh_2017thermodynamic,
  title={Thermodynamic equilibration of the carbon vacancy in 4H-SiC: A lifetime limiting defect},
  author={Ayedh, HM and Nipoti, R and Hall{\'e}n, Anders and Svensson, BG},
  journal={Journal of Applied Physics},
  volume={122},
  number={2},
  year={2017},
  publisher={AIP Publishing}
}

@Article{Emery_1965,
  author    = {F. E. Emery and T. A. Rabson},
  title     = {Average {E}nergy {E}xpended {P}er {I}onized {E}lectron-{H}ole {P}air in {S}ilicon and {G}ermanium as a {F}unction of {T}emperature},
  journal   = {Physical Review},
  year      = {1965},
  volume    = {140},
  number    = {6A},
  pages     = {A2089--A2093},
  month     = dec,
  doi       = {10.1103/physrev.140.a2089},
  publisher = {American Physical Society ({APS})},
  url       = {https://journals.aps.org/pr/abstract/10.1103/PhysRev.140.A2089},
}

@Book{Choyke_2004,
  title     = {Silicon carbide: recent major advances},
  publisher = {Springer},
  year      = {2004},
  author    = {Choyke, W. J. and Matsunami, H. and Pensl, G.},
}

@Article{Ivanov_2005,
  author    = {I. G. Ivanov and A. Henry and E. Janz{\'{e}}n},
  title     = {Ionization energies of phosphorus and nitrogen donors and aluminum acceptors in {4H} silicon carbide from the donor-acceptor pair emission},
  journal   = {Physical Review B},
  year      = {2005},
  volume    = {71},
  number    = {24},
  month     = jun,
  doi       = {10.1103/physrevb.71.241201},
  publisher = {American Physical Society ({APS})},
}

@Article{Prokscha_2012,
  author    = {T. Prokscha},
  title     = {Monte-{C}arlo {S}imulation of {T}ransitions between {D}ifferent {M}uonium {S}tates},
  journal   = {Physics Procedia},
  year      = {2012},
  volume    = {30},
  pages     = {50--54},
  doi       = {10.1016/j.phpro.2012.04.038},
  publisher = {Elsevier {BV}},
}

@Article{Morenzoni_2000,
  title={Low-energy $\mu$SR at PSI: present and future},
  author={Morenzoni, E and Gl{\"u}ckler, H and Prokscha, T and Weber, HP and Forgan, EM and Jackson, TJ and Luetkens, H and Niedermayer, Ch and Pleines, M and Birke, M and others},
  journal={Physica B: Condensed Matter},
  volume={289},
  pages={653--657},
  year={2000},
  publisher={Elsevier}
}

@Article{Woerle_2019a,
  title={Interaction of low-energy muons with defect profiles in proton-irradiated {Si} and {4H-SiC}},
  author={Woerle, Judith and Prokscha, Thomas and Hall{\'e}n, Anders and Grossner, Ulrike},
  journal={Physical Review B},
  volume={100},
  number={11},
  pages={115202},
  year={2019},
  publisher={APS}
}

@ARTICLE{Bathen_2020,
  author = {M. E. Bathen and A. Galeckas and J. Coutinho and L. Vines},
  title = {Influence of hydrogen implantation on emission from the silicon vacancy	in 4H-SiC},
  journal = {Journal of Applied Physics},
  year = {2020},
  volume = {127},
  pages = {085701},
  doi = {10.1063/1.5140659}
}

@article{Bathen2024Cinj,
  title={Impact of carbon injection in 4H-SiC on defect formation and minority carrier lifetime},
  author={Bathen, Marianne Etzelm{\"u}ller and Karsthof, Robert and Galeckas, Augustinas and Kumar, Piyush and Kuznetsov, Andrej Yu and Grossner, Ulrike and Vines, Lasse},
  journal={Materials Science in Semiconductor Processing},
  volume={176},
  pages={108316},
  year={2024},
  publisher={Elsevier}
}

@article{BaniSalameh_2007,
  title={Charge-state transitions of muonium in 6H silicon carbide},
  author={Bani-Salameh, HN and Meyer, AG and Carroll, BR and Lichti, RL and Celebi, YG and Chow, KH and King, PJC and Cox, SFJ},
  journal={Physica B: Condensed Matter},
  volume={401},
  pages={631--634},
  year={2007},
  publisher={Elsevier}
}

@inproceedings{Blanque_2004,
  title={Room temperature implantation and activation kinetics of nitrogen and phosphorus in 4H-SiC crystals},
  author={Blanqu{\'e}, Servane and P{\'e}rez, R and Godignon, Philippe and Mestres, Narcis and Morvan, Erwan and Kerlain, Alexandre and Dua, Christian and Brylinski, Christian and Zielinski, Marcin and Camassel, Jean},
  booktitle={Materials Science Forum},
  volume={457},
  pages={893--896},
  year={2004},
  organization={Trans Tech Publ}
}

@article{Karsthof_2022Cint,
    author = {Karsthof, Robert and Etzelmüller Bathen, Marianne and Kuznetsov, Andrej and Vines, Lasse},
    title = "{Formation of carbon interstitial-related defect levels by thermal injection of carbon into n-type 4H-SiC}",
    journal = {Journal of Applied Physics},
    volume = {131},
    number = {3},
    pages = {035702},
    year = {2022},
    month = {01},
    issn = {0021-8979},
    doi = {10.1063/5.0077308}
}

@ARTICLE{Kaukonen_2003,
  author = {M. Kaukonen and C. J. Fall and J. Lento},
  title = {Interstitial H and H2 in SiC},
  journal = {Applied Physics Letters},
  year = {2003},
  volume = {83},
  pages = {923},
  doi = {10.1063/1.1598646}
}

@Article{Kimoto_1995a,
  author  = {Kimoto,T. and Itoh,A. and Matsunami,H. and Sridhara,S. and Clemen,L. L. and Devaty,R. P. and Choyke,W. J. and Dalibor,T. and Pepperm{\"u}ller,C. and Pensl,G.},
  journal = {Applied Physics Letters},
  title   = {Nitrogen donors and deep levels in high-quality 4H-SiC epilayers grown by chemical vapor deposition},
  year    = {1995},
  number  = {19},
  pages   = {2833-2835},
  volume  = {67},
  doi     = {10.1063/1.114800},
  eprint  = {https://doi.org/10.1063/1.114800},
}

@article{Linnarsson_2018,
	title = {Surface {Erosion} of {Ion}-{Implanted} {4H}-{SiC} during {Annealing} with {Carbon} {Cap}},
	volume = {924},
	issn = {1662-9752},
	url = {https://www.scientific.net/MSF.924.373},
	doi = {10.4028/www.scientific.net/MSF.924.373},
	language = {en},
	urldate = {2024-11-26},
	journal = {Materials Science Forum},
	author = {Linnarsson, Margareta K. and Ayedh, Hussein M. and Hallén, Anders and Vines, Lasse and Svensson, Bengt Gunnar},
	month = jun,
	year = {2018},
	pages = {373--376},
}

@article{Prokscha_2020,
  title={Direct observation of hole carrier-density profiles and their light-induced manipulation at the surface of Ge},
  author={Prokscha, T and Chow, K-H and Salman, Z and Stilp, E and Suter, A},
  journal={Physical Review Applied},
  volume={14},
  number={1},
  pages={014098},
  year={2020},
  publisher={APS}
}

@Article{Simoes_2020,
  author  = {Sim{\~o}es,A. F. A. and Alberto,H. V. and Vil{\~a}o,R. C. and Gil,J. M. and Cunha,J. M. V. and Curado,M. A. and Salomé,P. M. P. and Prokscha,T. and Suter,A. and Salman,Z.},
  journal = {Review of Scientific Instruments},
  title   = {Muon implantation experiments in films: Obtaining depth-resolved information},
  year    = {2020},
  number  = {2},
  pages   = {023906},
  volume  = {91},
  doi     = {10.1063/1.5126529},
}

@article{Vassilevski_2005,
year = {2005},
month = {feb},
publisher = {},
volume = {20},
number = {3},
pages = {271},
author = {K V Vassilevski and N G Wright and I P Nikitina and A B Horsfall and A G O'Neill and M J Uren and K P Hilton and A G Masterton and A J Hydes and C M Johnson},
title = {Protection of selectively implanted and patterned silicon carbide surfaces with graphite capping layer during post-implantation annealing},
journal = {Semiconductor Science and Technology}
}

@ARTICLE{Vilao_2017,
  author = {R. C. Vil{\~a}o and R. B. L. Vieira and H. V. Alberto and J. M. Gil
	and A. Weidinger},
  title = {Role of the transition state in muon implantation},
  journal = {Physical Review B},
  year = {2017},
  volume = {96},
  pages = {195205},
  doi = {10.1103/PhysRevB.96.195205}
}

@inbook{Ziegler_1985,
  author       = {Ziegler, James F. and Biersack, Jochen P.},
  title        = {The Stopping and Range of Ions in Matter},
  booktitle    = {Treatise on Heavy-Ion Science: Volume 6: Astrophysics, Chemistry, and Condensed Matter},
  chapter      = {3},
  pages        = {93--129},
  year         = {1985},
  publisher    = {Springer US},
  address      = {Boston, MA},
  isbn         = {978-1-4615-8103-1},
  doi          = {10.1007/978-1-4615-8103-1_3},
  url          = {https://doi.org/10.1007/978-1-4615-8103-1_3}
}

@Article{Ziegler_2010,
  author    = {J. F. Ziegler and M.D. Ziegler and J.P. Biersack},
  journal   = {Nuclear Instruments and Methods in Physics Research Section B: Beam Interactions with Materials and Atoms},
  title     = {{SRIM} - {T}he stopping and range of ions in matter},
  year      = {2010},
  month     = jun,
  number    = {11-12},
  pages     = {1818--1823},
  volume    = {268},
  doi       = {10.1016/j.nimb.2010.02.091},
  publisher = {Elsevier {BV}},
  url       = {https://doi.org/10.1016/j.nimb.2010.02.091},
}

@Article{Woerle_2020,
  title={Muon Interaction with Negative-U and High-Spin-State Defects: Differentiating Between {C} and {Si} Vacancies in {4H-SiC}},
  author={Woerle, Judith and Bathen, Marianne Etzelm{\"u}ller and Prokscha, Thomas and Galeckas, Augustinas and Ayedh, Hussein M and Vines, Lasse and Grossner, Ulrike},
  journal={Physical Review Applied},
  volume={14},
  number={5},
  pages={054053},
  year={2020},
  publisher={APS}
}

@Article{Alberto_2018a,
  author    = {H. V. Alberto and R. C. Vil{\~{a}}o and R. B. L. Vieira and J. M. Gil and A. Weidinger and M. G. Sousa and J. P. Teixeira and A. F. da Cunha and J. P. Leit{\~{a}}o and P. M. P. Salom{\'{e}} and P. A. Fernandes and T. T{\"{o}}rndahl and T. Prokscha and A. Suter and Z. Salman},
  journal   = {Physical Review Materials},
  title     = {Slow-muon study of quaternary solar-cell materials: {S}ingle layers and p-n junctions},
  year      = {2018},
  month     = {feb},
  number    = {2},
  pages     = {025402},
  volume    = {2},
  doi       = {10.1103/physrevmaterials.2.025402},
  issue     = {2},
  numpages  = {11},
  publisher = {American Physical Society ({APS})},
}

@article{Hitti_1999,
  title = {Dynamics of negative muonium in n-type silicon},
  author = {Hitti, B. and Kreitzman, S. R. and Estle, T. L. and Bates, E. S. and Dawdy, M. R. and Head, T. L. and Lichti, R. L.},
  journal = {Phys. Rev. B},
  volume = {59},
  issue = {7},
  pages = {4918--4924},
  numpages = {0},
  year = {1999},
  month = {Feb},
  publisher = {American Physical Society}
}

@article{holzschuh_direct_1981,
	title = {Direct measurement of the muonium hyperfine frequencies in quartz},
	url = {https://www.e-periodica.ch/digbib/view?pid=hpa-001:1981:54::710},
	author = {Holzschuh, E. and Kündig, W. and Patterson, B. D.},
    journal = {Helvetica Physica Acta},
	volume={54},
  number={4},
  pages={609--609},
  year={1981},
}

@book{Blundell_2021_muon_spectroscopy,
	title = {{Muon Spectroscopy - An Introduction}},
    publisher = {{Oxford University Press}},
    address = {Oxford},
    editor = {Stephen J. Blundell and Roberto De Renzi and Tom Lancaster and Francis L. Pratt},
	year = {2021}
}

@article{Hillier_2022_primer,
title={Muon spin spectroscopy},
  author={Hillier, Adrian D and Blundell, Stephen J and McKenzie, Iain and Umegaki, Izumi and Shu, Lei and Wright, Joseph A and Prokscha, Thomas and Bert, Fabrice and Shimomura, Koichiro and Berlie, Adam and others},
  journal={Nature Reviews Methods Primers},
  volume={2},
  number={1},
  pages={1--24},
  year={2022},
  publisher={Nature Publishing Group}
}

@article{Kawahara_2010,
  title={Reduction of deep levels generated by ion implantation into n-and p-type 4H--SiC},
  author={Kawahara, Koutarou and Suda, Jun and Pensl, Gerhard and Kimoto, Tsunenobu},
  journal={Journal of Applied Physics},
  volume={108},
  number={3},
  pages={033706},
  year={2010},
  publisher={American Institute of Physics}
}

@article{Huang_2012,
doi = {10.1143/APEX.5.105802},
url = {https://dx.doi.org/10.1143/APEX.5.105802},
year = {2012},
month = {oct},
publisher = {},
volume = {5},
number = {10},
pages = {105802},
author = {Wei Huang and Shao-Hui Chang and Xue-Chao Liu and Biao Shi and Tian-Yu Zhou and Xi Liu and Cheng-Feng Yan and Yan-Qing Zheng and Jian-Hua Yang and Er-Wei Shi and Wen-Hua Zhang and Jun-Fa Zhu},
title = {Direct Observation of Nanoscale Native Oxide on 6H-SiC Surface and Its Effect on the Surface Band Bending},
journal = {Applied Physics Express}
}

@article{daSilva_2004,
  title={Crystalline silicon oxycarbide: Is there a native oxide for silicon carbide?},
  author={da Silva, Cesar RS and Justo, Jo{\~a}o F and Pereyra, In{\'e}s},
  journal={Applied physics letters},
  volume={84},
  number={24},
  pages={4845--4847},
  year={2004},
  publisher={American Institute of Physics}
}

@article{Das_2024,
  title={Study of Dopant Activation and Ionization for Phosphorus in 4H-SiC},
  author={Das, Suman and Lichtenwalner, Daniel J and Dixit, Hemant and Rogers, Steven and Scholze, Andreas and Ryu, Sei-Hyung},
  journal={Journal of Electronic Materials},
  volume={53},
  number={6},
  pages={2806--2810},
  year={2024},
  publisher={Springer}
}

@Article{Roccaforte_2021,
AUTHOR = {Roccaforte, Fabrizio and Fiorenza, Patrick and Vivona, Marilena and Greco, Giuseppe and Giannazzo, Filippo},
TITLE = {Selective Doping in Silicon Carbide Power Devices},
JOURNAL = {Materials},
VOLUME = {14},
YEAR = {2021},
NUMBER = {14},
ARTICLE-NUMBER = {3923},
PubMedID = {34300845},
ISSN = {1996-1944},
DOI = {10.3390/ma14143923}
}

@article{Pensl_2003,
  title={Implantation-induced defects in silicon carbide},
  author={Pensl, G and Frank, T and Krieger, M and Laube, M and Reshanov, S and Schmid, F and Weidner, M},
  journal={Physica B: Condensed Matter},
  volume={340},
  pages={121--127},
  year={2003},
  publisher={Elsevier}
}

@article{kuznetsov_dynamic_2003,
	title = {Dynamic annealing in ion implanted {SiC}: {Flux} versus temperature dependence},
	volume = {94},
	shorttitle = {Dynamic annealing in ion implanted {SiC}},
	url = {https://doi.org/10.1063/1.1622797},
	doi = {10.1063/1.1622797},
	journal = {Journal of Applied Physics},
	author = {Kuznetsov, A. Yu. and Wong-Leung, J. and Hallén, A. and Jagadish, C. and Svensson, B. G.},
	year = {2003},
	pages = {7112--7115},
}

@article{martins_depth_2023,
	title = {Depth profiling of {LE}-µ{SR} parameters with musrfit},
	volume = {2462},
	url = {https://dx.doi.org/10.1088/1742-6596/2462/1/012025},
	doi = {10.1088/1742-6596/2462/1/012025},
	urldate = {2023-04-15},
	journal = {Journal of Physics: Conference Series},
	author = {Martins, Maria Mendes and Suter, Andreas and Salman, Zaher and Prokscha, Thomas},
	year = {2023},
	pages = {012025},
}

@incollection{Goldberg2001SiC,
  author    = {Yu. Goldberg and M. E. Levinshtein and S. L. Rumyantsev},
  title     = {Silicon Carbide (SiC)},
  booktitle = {Properties of Advanced Semiconductor Materials: GaN, AlN, InN, BN, SiC, SiGe},
  editor    = {M. E. Levinshtein and S. L. Rumyantsev and M. S. Shur},
  publisher = {John Wiley \& Sons},
  address   = {New York},
  year      = {2001},
  pages     = {93--148}
}

\end{document}